\documentclass[12pt]{article}

\usepackage{graphicx}
\usepackage{cite}

\begin{document}

%%-move to normal A4-%%
\hoffset = -1truecm \voffset = -2truecm \baselineskip = 6 mm

\title{\bf Nucleon spin structure}

\author{
 {\bf Wei Zhu} and {\bf Jianhong Ruan}\\
\\
\normalsize Department of Physics, East China Normal University,
Shanghai 200062, P.R. China \\}

\date{}

\newpage

\maketitle

\vskip 3truecm

\begin{abstract}

     This paper contains three parts relating to the nucleon spin structure
in a simple picture of the nucleon: (i) The polarized
gluon distribution in the proton is dynamically predicted starting
from a low scale by using a nonlinear QCD evolution equation-the
DGLAP equation with the parton recombination corrections, where the
nucleon is almost only consisted of valence quarks. We find that the
contribution of the gluon polarization to the nucleon spin structure
is much larger than the predictions of most other theories. This
result suggests a significant orbital angular momentum of the gluons
is required to balance the gluon spin momentum; (ii) The spin
structure function $g_1^p$ of the proton is studied, where the
perturbative evolution of parton distributions and nonperturbative
Vector Meson Dominance (VMD) model are used. We predict $g_1^p$
asymptotic  behavior at small $x$ from lower $Q^2$ to higher $Q^2$.
The results are compatible with the data including the HERA early
estimations and COMPASS new results; (iii) The generalized
Gerasimov-Drell-Hearn (GDH) sum rule is understood based on the
polarized parton distributions of the proton with the higher twist
contributions. A simple parameterized formula is proposed to clearly
present the contributions of different components in the proton to
$\Gamma_1^p(Q^2)$. The results suggest a possible extended objects
with size $0.2-0.3~ fm$ inside the proton.

\end{abstract}

PACS number(s): 12.38.Cy, 12.38.Qk, 12.38.Lg, 12.40.Vv

$keywords$: Nucleon spin structure

\newpage
\begin{center}
\section{Introduction}
\end{center}

    A precise determination of the polarized gluon distribution
$\delta g(x,Q^2)$ is important in order to understand the spin
structure of the nucleon. However, the direct measurement of the
polarized gluon distribution in the nucleon is difficult. In the
global analysis of the polarized lepton-nucleon deep inelastic
scattering (DIS), the distribution $\delta g(x,Q^2)$ is extracted
from the spin structure function $g_1(x,Q^2)$ through scaling
violation as a higher-order effect of quantum chromodynamics (QCD).
Unfortunately, such indirect determination of $\delta g(x,Q^2)$ is
affected by large uncertainties because of the limited range in
momentum transfer at fixed Bjorken-$x$ and almost entirely arbitrary
input gluon distribution. In fact, the data suggest that such global
fit with either positive or negative input gluon distributions
provides equally good agreement. Recently, a high-precision
measurement of the mid-rapidity polarized proton-proton (p-p) collisions stringently
constraint the polarized parton distribution functions mentioned
above. The analysis of NNPDF collaboration [1] found an evidence for
possible larger gluon spin distribution, which is against the common
belief that it is rather small.

    In Sec. 2 of this work we use a QCD dynamic model of the parton distributions to predict
the polarized gluon distribution in the proton without unknown input
gluon distribution. Our model imagines that all gluons in the nucleon
are radiated from the intrinsic quarks beginning at a low resolution
scale. Thus, we can predict the radiative (unpolarized and
polarized) gluon distributions provided the initial quark
distributions are fixed. Such quark model was early proposed by [2,3,4]
in 1977, it was improved in our previous work [5,6], where the
Dokshitzer-Gribov-Lipatov-Altarelli-Parisi (DGLAP) equation [7,8,9] with
the parton recombination corrections is used to reproduce the
unpolarized parton distributions of the nucleon at $Q^2>1 GeV^2$
[10,11,12]. Since the similar corrections of the parton recombination to
the polarized DGLAP equation have been proposed in work [13], we can
use these two modified evolution equations to  predict
the polarized gluon distribution in the proton dynamically.

    Comparing with the global analysis via scaling violation, the polarized gluon distribution in this
work is determined directly by the observed spin structure function
$g_1(x,Q^2)$. We find that the contribution of the gluon
polarization to the nucleon spin structure is surprisedly large,
which is in excess of the previous estimations in theory. The
reasons are as following: (i) The shadowing effect of the gluon
recombination in the evolution of the polarized parton distributions
is weaker than that in the unpolarized case since $g\delta g<< g^2$
at small $x$. Therefore, much more strong polarized gluons are
emitted by quarks inside the polarized proton through a long
evolution length from $\mu^2$ to $Q^2> 1 GeV^2$; (ii) The positive
contribution of the polarized gluon recombination, which is opposite
to that of the unpolarized gluon recombination enhance the
accumulation of the gluon helicity. The QCD evolution of the parton
distributions begins from a low bound state scale $\mu^2$ not only
dynamically determine the polarized gluon distribution, but also
exposes a novel spin-orbital structure of the nucleon in the
light-cone frame, where the nucleon spin crisis has a possible
explanation.

    Concerning the spin structure function, recently COMPASS experiment
at CERN collected a large number of events of polarized inelastic
scattering off the protons with very small $x$ [14]. The preliminary
analysis of these data combining with the previous experiments [15,16,17,18],
showed non zero and positive asymptotic structure function $g_1^p$.
In these fixed target experiments the low values of $x$ are almost
reached by lowering the values of $Q^2$. The knowledge of the
nucleon spin structure function $g_1(x,Q^2)$ at low $Q^2$ and small
$x$ is particulary interesting, since it is not only an important
information to resolve the "proton spin crisis", but also provides
us with a good place to study the transition from the perturbative
research to the nonperturbative description of the proton structure.

    In Sec. 3 we try to study the behavior of $g_1^p$ at small $x$
but in the full $Q^2$ range. As we know that the structure functions
of the nucleon are mainly constructed of the parton distributions at
$Q^2>1 GeV^2$, while the nonperturbative contributions to the
structure functions become un-negligible at $Q^2\ll1~GeV^2$. A key
question is what components construct the spin structure functions
of the proton at such low $Q^2$? Particularly, do the parton
distributions and their pQCD evolution still play a role or not? To
answere these questions, we discuss the application of the DGLAP
equation with the parton recombination corrections at low $Q^2$ in
detail. We point out that the isolation of the contributions of the
vector meson is necessary for keeping the factorization schema of
the polarized parton distributions at low $Q^2$. We find two
different asymptotic behaviors of $g_1^p$ at $x<10^{-3}$:
nonperturbaive behavior  $\sim x^{-1}$ at $Q^2<1 GeV^2$ and
perturbative drop at $Q^2>3 GeV^2$. We predict the translation of
$g_1^p$ at small $x$ from lower $Q^2$ to higher $Q^2$. The results
are compatible with the data including the early HERA estimations
and COMPASS new results. We point out that the measurements at
different $x$ with different values of $Q^2$ in the fixed target
experiments mix the complicated asymptotic behavior of $g_1^p$. The
predicted strong $Q^2$- and $x$-dependence of $g_1^p$ at $0.01<Q^2<3
GeV^2$ and $x<0.1$ due to the mixture of nonperturbative vector
meson interactions and the QCD evolution of the parton distributions
can be checked on the next Electron-Ion Collider (EIC).

       There is particular interest in the first moment $\Gamma_1(Q^2)=\int_0^1 dxg_1(x,Q^2)$
of the spin structure functions $g_1(x,Q^2)$, which has been
measured from high $Q^2$ down to $\sim 0~ GeV^2$. The goal to obtain
universal expressions describing $\Gamma_1(Q^2)$ at any $Q^2$ is an
attractive task for both theoretical and phenomenological point of
view. In theory, $\Gamma_1(0)$ is constrained by the
Gerasimov-Drell-Hearn (GDH) sum rule [19,20]. In Sec. 4 we try to
expose the partonic structure in the GDH sum rule. Since we have
known the contributions of $g_1^{DGLAP+ZRS}$ and $g_1^{VMD}$, one
can expose the properties of $\Gamma_1^{HT}(Q^2)$ after subtracting
these two contributions from the experimental data about
$\Gamma_1^p(Q^2)$. This opens a window to visit higher twist effects
at low $Q^2$ in the nucleon structure. We proposed  a simple
parameterized form of $\Gamma_1^p(Q^2)$. We find that the negative
twist-4 effect dominates the suppression of $\Gamma_1^p(Q^2)$ at
$Q^2<1GeV^2$, while both the twist-4 and twist-6 effects have a
dramatic change of $\Gamma_1^p(Q^2)$ at $Q^2\sim 1 GeV^2$, which
suggest a possible extended objects with size $0.2-0.3~ fm$ inside
the proton.

    Finally, following the above mentioned discussions, we will
give a summary in Sec.5.

\newpage
\begin{center}
\section { Dynamical determination of gluon helicity distribution
in the nucleon}

\subsection{ Nonlinear polarized QCD evolution equation}
\end{center}

      We use $f_+(x,Q^2)$ and $f_-(x,Q^2)$ to refer to parton ($f=q,\overline{q},g$) densities
with positive and negative helicity which carry a fraction $x$ of
the nucleon momentum. The difference $\delta f(x,Q^2) =
f_+(x,Q^2)-f_-(x,Q^2)$ measures how much the parton of flavor f
remembers its parent's nucleon polarization. The spin
averaged parton densities are given by $f(x,Q^2) =
f_+(x,Q^2)+f_-(x,Q^2)$.

     The spin-dependent QCD evolution equation of
parton distributions with parton recombination corrections was first
derived by Zhu, Shen and Ruan (ZRS) in [13], it reads

$$Q^2\frac{dx\delta q_v(x,Q^2)}{dQ^2}$$
$$=\frac{\alpha_s(Q^2)}{2\pi}\int_x^1 \frac{dy}{y}\frac{x}{y}y\delta q_v(y,Q^2)\Delta P_{qq}(\frac{x}{y}), \eqno(2.1.1)$$
for flavor non-singlet quarks;
$$Q^2\frac{dx\delta q_i(x,Q^2)}{dQ^2}$$
$$=\frac{\alpha_s(Q^2)}{2\pi}\int_x^1\frac{dy}{y}\frac{x}{y}[y \delta q_i(y,Q^2)\Delta P_{qq}(\frac{x}{y})+y\delta g(y,Q^2)\Delta P_{qg}(\frac{x}{y})]$$
$$-\frac{\alpha_s^2(Q^2)}{4\pi R^2 Q^2}\int_x^{1/2}\frac{dy}{y}x
\Delta P_{gg\rightarrow q}(x,y)[yg(y,Q^2)y\delta g(y,Q^2)]$$
$$+\frac{\alpha_s^2(Q^2)}{4\pi R^2 Q^2}\int_{x/2}^x\frac{dy}{y}x
\Delta P_{gg\rightarrow q}(x,y)[yg(y,Q^2)y\delta g(y,Q^2)], (if~x\le
1/2),$$

$$Q^2\frac{dx\delta q_i(x,Q^2)}{dQ^2}$$
$$=\frac{\alpha_s(Q^2)}{2\pi}\int_x^1\frac{dy}{y}\frac{x}{y}[y
\delta q_i(y,Q^2)\Delta P_{qq}(\frac{x}{y})+y\delta g(y,Q^2) \Delta
P_{qg}(\frac{x}{y})]$$
$$+\frac{\alpha_s^2(Q^2)}{4\pi R^2 Q^2}\int_{x/2}^{1/2}\frac{dy}{y}x
\Delta P_{gg\rightarrow q}(x,y)[yg(y,Q^2)y\delta
g(y,Q^2)],(if~1/2\le x\le 1),\eqno(2.1.2)$$ for sea quarks;

$$Q^2\frac{dx\delta g(x,Q^2)}{dQ^2}$$
$$=\frac{\alpha_s(Q^2)}{2\pi}\int_x^1\frac{dy}{y}\frac{x}{y}[y  \sum_{i=1}^{2f}\delta q_i(y,Q^2)\Delta P_{qq}(\frac{x}{y})+y\delta g(y,Q^2) \Delta P_{gg}(\frac{x}{y})]$$
$$-\frac{\alpha_s^2(Q^2)}{4\pi R^2 Q^2}\int_x^{1/2}\frac{dy}{y}x
\Delta P_{gg\rightarrow g}(x,y)[ yg(y,Q^2)y\delta g(y,Q^2)]$$
$$+\frac{\alpha_s^2(Q^2)}{4\pi R^2 Q^2}\int_{x/2}^x\frac{dy}{y}x
\Delta P_{gg\rightarrow g}(x,y)[yg(y,Q^2)y\delta g(y,Q^2)],(if~x\le
1/2),$$

$$Q^2\frac{dx\delta g(x,Q^2)}{dQ^2}$$
$$=\frac{\alpha_s(Q^2)}{2\pi}\int_x^1\frac{dy}{y}\frac{x}{y}[y\sum_{i=1}^{2f}\delta q_i(y,Q^2)\Delta P_{qq}(\frac{x}{y})+y\delta g(y,Q^2)\Delta P_{gg}(\frac{x}{y})]$$
$$+\frac{\alpha_s^2(Q^2)}{4\pi R^2 Q^2}\int_{x/2}^{1/2}\frac{dy}{y}x
\Delta P_{gg\rightarrow g}(x,y)[ yg(y,Q^2)y\delta
g(y,Q^2)],(if~1/2\le x\le 1)\eqno(2.1.3)$$ for gluon, where the
factor $1/(4\pi R^2)$ is from the normalization of two-parton
distribution, R is the correlation length of two initial partons,
the linear terms are the standard spin-dependent DGLAP evolution and
the recombination functions in the nonlinear terms are(see appendix A)

$$\Delta P_{gg\rightarrow
g}(x,y)=\frac {27}{64}\frac
{(2y-x)(-20y^3+12y^2x-x^3)}{y^5},\eqno(2.1.4)$$

$$\Delta P_{gg\rightarrow q}(x,y)=\frac {1}{48}\frac {(2y-x)^2
(-y+x)}{y^4}.\eqno(2.1.5)$$

    The spin structure function $g_1$ at leading order (LO) and
$Q^2>1GeV^2$ is written as

$$g_1(x,Q^2)=\frac{1}{2}\sum_i e_i^2[\delta q_i(x,Q^2)+\delta \overline{q}_i(x,Q^2)], \eqno(2.1.6)$$
where $e_i$ is the electric charge of the (light) quark of flavor i,
$i = u, d, s$.

    The solutions of Eqs (2.1.1-2.1.3) are coupled with the
spin-averaged evolution equations, which are

$$Q^2\frac{dx q_v(x,Q^2)}{dQ^2}$$
$$=\frac{\alpha_s(Q^2)}{2\pi}\int_x^1 \frac{dy}{y}\frac{x}{y}y q_v(y,Q^2) P_{qq}(\frac{x}{y}), \eqno(2.1.7)$$
for valence quarks;

$$Q^2\frac{dxq_i(x,Q^2)}{dQ^2}$$
$$=\frac{\alpha_s(Q^2)}{2\pi}\int_x^1\frac{dy}{y}\frac{x}{y}[yq_i(y,Q^2)P_{qq}(\frac{x}{y})+yg(y,Q^2)P_{qg}(\frac{x}{y})]$$
$$-\frac{\alpha_s^2(Q^2)}{4\pi R^2 Q^2}\int_x^{1/2}\frac{dy}{y}x
P_{gg\rightarrow q}(x,y)[ yg(y,Q^2)]^2$$
$$+\frac{\alpha_s^2(Q^2)}{4\pi R^2 Q^2}\int_{x/2}^x\frac{dy}{y}x
P_{gg\rightarrow q}(x,y)[ yg(y,Q^2)]^2, (if~x\le 1/2),$$

$$Q^2\frac{dxq_i(x,Q^2)}{dQ^2}$$
$$=\frac{\alpha_s(Q^2)}{2\pi}\int_x^1\frac{dy}{y}\frac{x}{y}[yq_i(y,Q^2)P_{qq}(\frac{x}{y})+yg(y,Q^2)P_{qg}(\frac{x}{y})]$$
$$+\frac{\alpha_s^2(Q^2)}{4\pi R^2 Q^2}\int_{x/2}^{1/2}\frac{dy}{y}x
P_{gg\rightarrow q}(x,y)[ yg(y,Q^2)]^2,(if~1/2\le x\le
1),\eqno(2.1.8)$$ for sea quarks;

$$Q^2\frac{dxg(x,Q^2)}{dQ^2}$$
$$=\frac{\alpha_s(Q^2)}{2\pi}\int_x^1\frac{dy}{y}\frac{x}{y}[y  \sum_{i=1}^{2f}q_i(y,Q^2)P_{qq}(\frac{x}{y})+yg(y,Q^2) P_{gg}(\frac{x}{y})]$$
$$-\frac{\alpha_s^2(Q^2)}{4\pi R^2 Q^2}\int_x^{1/2}\frac{dy}{y}x
P_{gg\rightarrow g}(x,y)[ yg(y,Q^2)]^2$$
$$+\frac{\alpha_s^2(Q^2)}{4\pi R^2 Q^2}\int_{x/2}^x\frac{dy}{y}x
P_{gg\rightarrow g}(x,y)[yg(y,Q^2)]^2,(if~x\le 1/2),$$

$$Q^2\frac{dxg(x,Q^2)}{dQ^2}$$
$$=\frac{\alpha_s(Q^2)}{2\pi}\int_x^1\frac{dy}{y}\frac{x}{y}[y  \sum_{i=1}^{2f} q_i(y,Q^2) P_{qq}(\frac{x}{y})+y g(y,Q^2) P_{gg}(\frac{x}{y})]$$
$$+\frac{\alpha_s^2(Q^2)}{4\pi R^2 Q^2}\int_{x/2}^{1/2}\frac{dy}{y}x
P_{gg\rightarrow g}(x,y)[ yg(y,Q^2)]^2,(if~1/2\le x\le
1)\eqno(2.1.9)$$ for gluon, where the linear terms are the standard
DGLAP evolution [7,8,9] and the recombination functions in the nonlinear
terms are [10,11,12]

$$P_{gg\rightarrow
g}(x,y)=\frac{9}{64}\frac{(2y-x)(72y^4-48xy^3+140x^2y^2-116x^3y+29x^4)}{xy^5},\eqno(2.1.10)$$

$$P_{gg\rightarrow q}(x,y)=P_{gg\rightarrow \overline{q}}(x,y)=
\frac{1}{96}\frac{(2y-x)^2(18y^2-21xy+14x^2)}{y^5}.\eqno(2.1.11)$$

\newpage
\begin{center}
\subsection{Dynamically radiative polarized gluon distribution}
\end{center}

    We focus on the gluon distribution, its
evolution is dominated by the valence quark distributions.
Therefore, the corrections of the asymmetry sea distributions in the
nucleon are neglected in this work.

    As we know, there are many effective QCD theories which describe the nucleon
as a bound state of three quarks in its rest frame. The
distributions of these quarks in the light-cone configuration have a
similar character as the valence quark distributions observed at
high $Q^2$ in DIS. Besides, the QCD evolution equation shows that
either the second moment (i.e., the average momentum fraction) of
the unpolarized gluon distribution or the first moment (i.e., the
total helicity) of the polarized gluon distribution increase as
$Q^2$ increasing. A natural suggestion is that all partons (valence
quarks, sea quarks and gluons) at high $Q^2$ scale are evolved from
three initial valence quarks via  QCD dynamics. Such idea was
first proposed in 1977 for unpolarized parton distributions by
[2,3,4]. They assumed that the nucleon consists of valence quarks at a
low starting point $\mu^2\sim 0.064 GeV^2$ (but is still in the
perturbative region $\alpha_s(\mu^2)/2\pi <1$ and
$\mu>\Lambda_{QCD}$), and the gluons and sea quarks are produced at
$Q^2>\mu^2$ using the DGLAP equation. However, such natural input is
failed due to the too steep behavior of the predicted parton
distributions at small $x$ since a long evolution distance from
$\mu^2$ to $Q^2>1 GeV^2$. Recently the above naive idea was realized
in the unpolarized DGLAP equation with the parton recombination
corrections at LO approximation in [5,6], where the input
distributions at $\mu^2=0.064GeV^2$ were extracted through fitting
$F_2^{p,n}(x,Q^2)$ and they have been fixed as

$$
xu_v(x,\mu^2)=24.3x^{1.98}(1-x)^{2.06},\eqno(2.2.1)
$$

$$
xd_v(x,\mu^2)=9.10x^{1.31}(1-x)^{3.8}, \eqno(2.2.2)$$ while

$$g(x,\mu^2)=0,~~q_i(x,\mu^2)=\overline {q}_i(x,\mu^2)=0,
\eqno(2.2.3)$$and the parameters in Eqs. (2.1.8) and (2.1.9) are
$\Lambda_{QCD}=0.204 GeV$ and $R=4.24GeV^{-1}$. We plot these input
distributions in Fig.1.

    Similarly, using the polarized DGLAP equation with the parton recombination corrections Eqs. (2.1.1)-(2.1.3)
and combining Eqs. (2.1.7)-(2.1.9), we fit $g_1^{p,n}(x,Q^2)$ with
the data [21] in Fig.2, and extract the input polarized valence
quark distributions in the proton as

$$\delta u_v(x,\mu^2)=40.3x^{2.85}(1-x)^{2.15},\eqno(2.2.4)$$

$$\delta d_v(x,\mu^2)=-18.22x^{1.41}(1-x)^{4.0},\eqno(2.2.5)$$and

$$\delta g(x,\mu^2)=0,~~\delta q_i(x,\mu^2)=\delta \overline
{q}_i(x,\mu^2)=0,\eqno(2.2.6)$$they are plotted in Fig.1.

    We predict the polarized gluon distribution at different $Q^2$ in Fig. 3.
The results clearly show the accumulation of polarized gluons at
small $x$.

    There are several databases of the polarized parton distributions,
which are extracted by the global fitting DIS data. For example, we
compare our results with the GRV distribution [22] in Fig. 4. The
difference is obvious. It is not surprise that the polarized gluon
distribution has large uncertainty, since the shape of the input
gluon distribution is not constrained well enough by the DIS data
alone.

  In order to understand the contribution of large gluon polarization, we draw the evolution kernels
$P_{qg}(z),P_{gg}(z),$ $ \Delta P_{qg}(z),$ $\Delta P_{gg}(z)$ and
$yP_{gg\rightarrow g}(z), yP_{gg\rightarrow q}(z)$, $y\Delta
P_{gg\rightarrow g}(z), y\Delta P_{gg\rightarrow q}(z)$ in Figs. 5
and 6. One can find that

(i) $P_{gg}(z)>0$ and $\Delta P_{gg}(z)>0$ imply that $\delta
g(x,Q^2)$ is positive in our dynamic model;

(ii) Since $\Delta P_{qg}(z)<0$ at small $z$, we have
$dg_1(x,Q^2)/d\ln Q^2\sim -\delta g(x,Q^2)$ at small $x$, i.e., a
large positive $\delta g$ at small $x$ is expected to drive $g_1$
towards large negative values;

(iii) $\Delta P_{gg\rightarrow g}<0$ and $\Delta P_{gg\rightarrow
q}<0$ lead the net positive corrections of the gluon fusion to the
polarized parton distributions since a negative sign in the
shadowing terms of Eq. (2.1.3). To illustrate this effect, in Fig.7
we present $x\delta g(x,Q^2)$  with and without the corrections of
gluon recombination corrections at $Q^2=1$ and $5 GeV^2$. One can
find that the effects of the gluon recombination in the polarized
gluon distribution is positive.

     As we know that some approaches are planned to measure the
gluon distributions. For example, the semi-inclusive deep inelastic
scattering processes  measure the $\delta g/g$ from helicity
asymmetry in photon-gluon fusion. The COMPASS collaboration [23]
have used this method and found a rather small value for $\delta
g/g= 0.024\pm 0.080\pm 0.057$ at $x=0.09$ and $Q^2=3GeV^2$. However
we think that although the value of $\delta g/g$ is small, the
polarized gluon contribution to the spin of the nucleon may be
sizable since $g$ itself is large at small $x$. In order to compare
with the data, one needs to assume a suitable form for the
unpolarized gluon distribution $g(x,Q^2)$. Fortunately, both $\delta
g(x,Q^2)$ and $g(x,Q^2)$ are calculated within a same dynamics in
this work and we avoid a larger uncertainty in the determination of
$g(x,Q^2)$. We compare our predicted $\delta g/g$ with the COMPASS
data in Fig. 8.

    The other direct probing of $\delta g$ is offered by jet and $\pi$ production
in polarized proton-proton collisions available at BNL Relativistic
Heavy Ion Collider (RHIC). A recent DSSV analysis [24] of
high-statistics 2009 STAR [25] and PHENIX [26] data showed an
evidence of non-zero gluon helicity in the proton. They found that
the polarized gluon distribution in the proton is positive and away
from zero in $0.05<x<0.2$, although the presented data  has very
large uncertainty at small $x$ region. Figure 9 presents the
comparisons of our predicted $g_1^p(x,Q^2)$ at $Q^2=10 GeV^2$ with
the DSSV bounds. Our results are beyond a up bound of the DSSV
results, however, a sizable gluon polarization is still possible if
taking the $90\%$ confidence level (C.L.) interval.

    The NNPDF group has developed a new methodology
[1] to extract polarized gluon distribution function.
They used all essential available data and got an evidence of
positive gluon polarization in the medium and small $x$ region.
This discovery is compatible with our results. Figure 10 shows the
comparison of our predicted polarized gluon distribution with the
NNPDF bounds. This example shows that a positive initial
distribution of the polarized gluon at $\mu^2$ in the nucleon is
impossible since it will obviously go beyond the up bound of the NNPDF
analysis at $Q^2\sim 10GeV^2$.

\newpage
\begin{center}
\subsection{Discussions}
\end{center}

     The total helicity of partons in a polarized proton are calculated by
the first moments

$$\Delta f(Q^2)=\int_0^1dx\delta f(x,Q^2), \eqno(2.3.1)$$
Note that our predicted $\Delta q_s(Q^2)$ for the sea quarks is
positive since the negative contributions from the asymmetric
strange quarks are neglected in this work.

     The contribution of quark polarization to the proton spin $\Delta
\Sigma(Q^2=5 GeV^2)\simeq 0.30$ in our estimation. It is interesting
that this value is compatible with the world average values $\Delta
\Sigma(Q^2=10GeV^2)=0.31\pm 0.07$ [27], and $\Delta \Sigma(Q^2=5
GeV^2)=0.333\pm 0.011\pm0.025\pm 0.028$ [28].

    In Fig. 11 we plot the evolutions of
$\Delta\Sigma(Q^2)$ and $\Delta g(Q^2)$ with increasing $Q^2$. We
find that $\Delta q_v(Q^2)=0.296$, but $\Delta q_s(Q^2)$ is slowly
increasing from $0$ at $\mu^2$ to $0.016$ at $Q^2=1000GeV^2$ due to
the parton recombination corrections. On the other hand, the gluon
helicity $\Delta g(Q^2)$ increases with $\ln Q^2$  beginning from
zero at $\mu^2$, and becomes large when $Q^2>0.3 GeV^2$.

    The above mentioned $\Delta \Sigma$ and $\Delta g$ should be
balanced by the orbital angular momenta of partons. For this sake,
we write the nucleon helicity sum rule

$$\frac{1}{2}=\frac{1}{2}\Delta\Sigma(Q^2)+\Delta g(Q^2)+\sum_qL_q^z(Q^2)+L_g^z(Q^2), \eqno(2.3.2)$$
where $L_{q,g}^z$ denote the contributions of orbital angular
momenta of quarks and gluons. In our model the sum rule at
$Q^2=\mu^2$ is

$$\frac{1}{2}=\frac{1}{2}\Delta\Sigma(\mu^2)+\sum_qL_q^z(\mu^2),
\eqno(2.3.3)$$where

$$\Delta\Sigma(\mu^2)=\Delta u_v(\mu^2)+\Delta d_v(\mu^2),\eqno(2.3.4)$$

$$\sum_q L_q^z(\mu^2)=L_u^z(\mu^2)+L_d^z(\mu^2).\eqno(2.3.5)$$

    On the other hand, the helicity sum rule of the polarized proton in its rest
frame according to the constituent quark model is

$$\frac{1}{2}=\frac{1}{2}\sum_q\Delta q^c,\eqno(2.3.6)$$where $\Delta q^c$ denotes
the quark polarization in the quark model.

    Comparing Eq. (2.3.6) with Eq. (2.3.3), we assume

$$\frac{1}{2}\Delta
u^c=\frac{1}{2}\Delta u_v(\mu^2)+L_{u_v}^z(\mu^2), $$ and

$$\frac{1}{2}\Delta
d^c=\frac{1}{2}\Delta d_v(\mu^2)+L_{d_v}^z(\mu^2),
\eqno(2.3.7)$$where the motions of partons are independent.

     Taking the SU(6) symmetry in the proton rest frame, we have

$$\Delta u^c=\frac{4}{3},~~~~ \Delta d^c=-\frac{1}{3}.\eqno(2.3.8)$$
From Eqs. (2.2.5) and (2.2.6) we know

$$\Delta u_v(\mu^2)=0.644, \Delta d_v(\mu^2)=-0.348.\eqno(2.3.9)$$
Using Eq. (2.3.7) we obtain

$$L_u^z(\mu^2)=0.345, ~~~~L_d^z(\mu^2)=0.007,\eqno(2.3.10)$$ it implies
that a polarized proton at scale $\mu^2$ has two rotating u-quarks,
while the d-quark is located at the center of the proton since it
has almost zero-orbital momentum.

    According to SU(6) symmetry, the constituent
quark has zero angular momentum. However, according to Ref. [29,30,31,32] the
orbital angular momentum $L_q^z(\mu^2)$ in Eq. (2.3.7) may origin
from the transverse distribution of the constituent quark in the
rest frame due to the Melosh-Wigher rotation [33]. The Melosh-Wigher
rotation is a pure kinematic effect in the frame transformation, we
consider that this effect should keep the angular momentum
conservation, for say,

$$\frac{2}{3}\vec{s}^c_u=b\vec{s}_u+c\vec{L}_u,\eqno(2.3.11)$$
where $\vec{s}_u$ and $\vec{L}_u$ are the spin and orbital angular
momentum of a u-valence quark at $\mu^2$ in the light-cone frame of
the polarized proton; $2/3$ is from the $SU(6)$-distribution, the
values of $b$ and $c$ depend on the wave function of valence quarks
at $\mu^2$ [29,30,31,32]. Because the spin is an elemental physical quantity,
it is always has $\vert \vec{s}^c_u\vert=\vert\vec{s}_u\vert\equiv1/2$,
therefore, the orbital angular momentum $\vec{L}_u$ changes only the
direction of the spin from $\vec{s}^c_u$ (it is also the polarized
direction of the proton) to $\vec{s}_u$. Under these constraint
conditions, once the values of $b$ and $c$ are determined by the
wave function of valence quarks, the coupling angle between
$\vec{s}_u$ and $\vec{L}_u$ in Eq. (2.3.11) can be fixed (see Fig. 12),
and it leads to $\Delta u_v(\mu^2)<\Delta u^c$.

    We discuss the evolution of the sum rule (2.3.2) with $Q^2$. The evolution
equation for the quark and gluon orbital angular momenta at the
leading order approximation was derived by Ji, Tang and Hoodbhoy in
[34], it reads

$$\frac{d\sum_qL_q^z(t)}{dt}=\frac{\alpha_s(t)}{2\pi}[-\frac{4}{3}C_F\sum_qL_q^z(t)+\frac{n_f}{3}L_g^z(t)]+
\frac{\alpha_s(t)}{2\pi}[-\frac{2}{3}C_F\Sigma(t)+\frac{n_f}{3}\Delta
g(t)],$$

$$\frac{dL_g^z(t)}{dt}=\frac{\alpha_s(t)}{2\pi}[\frac{4}{3}C_F\sum_qL_q^z(t)-\frac{n_f}{3}L_g^z(t)]+
\frac{\alpha_s(t)}{2\pi}[-\frac{5}{6}C_F\Sigma(t)-\frac{11}{2}\Delta
g(t)],\eqno(2.3.12)$$ where $C_F=4/3$, $n_f$ is the number of active
quark flavors, $t=\ln (Q^2/\Lambda^2)$ and $t_0=\ln
(\mu^2/\Lambda^2)$. The solutions are

$$\sum_qL_q^z(Q^2)=-\frac{1}{2}\Delta \Sigma(Q^2)+\frac{1}{2}\frac{3n_f}{16+3n_f}
+(\frac{t}{t_0})^{-2(16+3n_f)/9\beta_0}[\sum_qL_q^z(\mu^2)+\frac{1}{2}\Delta\Sigma(\mu^2)-\frac{1}{2}\frac{3n_f}{16+3n_f}],$$and

$$L_g^z(Q^2)=-\Delta g(Q^2)+\frac{1}{2}\frac{16}{16+3n_f}+
(\frac{t}{t_0})^{-2(16+3n_f)/9\beta_0}[L_g^z(\mu^2)+\Delta
g(\mu^2)-\frac{1}{2}\frac{16}{16+3n_f}], \eqno(2.3.13) $$where
$\beta_0=11-2n/3$. Because of
$\Delta\Sigma(\mu^2)=0.296,L_g^z(\mu^2)=0,\Delta g(\mu^2)=0$, we can
fixed $\sum_qL_q^z(\mu^2)=1/2-\Delta\Sigma(\mu^2)/2=0.352$.

Table.  The contributions of various components to the proton spin
at different $Q^2$.
\begin{center}
\begin{tabular}{c|c|c|c|c}
%\toprule
    \hline
    $Q^2$     &  $0.064GeV^2$   &  $1GeV^2$ &  $10GeV^2$ &$100GeV^2$\\\hline

    $\frac{1}{2}\Delta\Sigma $        & 0.148& 0.149                   &       0.151         &  0.153          \\\hline
    $\sum_qL_q^z$                             &0.352 &  0.124                  &       0.096        &  0.080       \\\hline
    $\Delta g$                        & 0    &  1.056                  &       1.993       &  2.889       \\\hline
    $L_g^z$                             & 0    & -0.829                  &       -1.74       &  -2.622        \\\hline
    Total                             &  $\frac{1}{2}$  &$\frac{1}{2}$  &     $\frac{1}{2}$       &  $\frac{1}{2}$        \\\hline

%    \bottomrule
    \end{tabular}
    \end{center}

    In Fig. 11 we add the curves of $\sum_qL_q^z(Q^2)$ and $L_g^z(Q^2)$
with $n_f=3$. These results describe a following novel
spin-orbital picture of the proton in the light-cone frame: The
proton is mainly constructed by one d-valence quark and two
u-valence quarks, the d-quark is located at the center of the
proton, and the two u-quarks rotate with $\sum_qL_q^z(\mu^2)\simeq
0.35$ around the d-quark at a bound state scale, with the $Q^2$
increasing the valence quarks radiate gluons and then sea quarks
follow them. The former builds fast rotating glue cloud (see
Table ), but their rotating direction is opposite to the u-quarks.
Remind that the above mentioned possible orbital angular momentum of
the partons in the polarized proton which can be checked in the
experiments [35,36].

\newpage
\begin{center}
\section {Spin structure function $g_1^p$ at small $x$}

\subsection{A general consideration of the nucleon structure function
at low $Q^2$ }
\end{center}

    In the research of the nucleon structure functions at
the full kinematic region, an argued question is whether the parton
distributions and their perturbative QCD evolution can (even partly)
be applied to the low $Q^2$ range or the parton concept is suddenly
invalid at a critical value of $Q^2\leq 1GeV^2$?

    Let us begin from the parton model for the spin-dependent distribution, which
is written based on the Collins-Soper-Sterman (CSS) factorization
schema [37] at the collinear approximation and in the twist-2 level,

$$g_1(x,Q^2)=\int^1_0\frac{dy}{y}\sum_qC_q(x/y,Q^2/\mu_F)\delta q(y,\mu_F),\eqno(3.1.1)$$
which breaks up the spin structure function into two factors
associated with perturbative short-distance functions $C_a$ and
nonperturbative polarized parton distributions $\delta q$ at the
factorization scale $\mu_F$.

 Taking the lowest order of $C_q$

$$C_q(x/y,Q^2/\mu_F)=\frac{1}{2}e_q^2\delta(x/y-1)\delta(Q-\mu_F)+{\cal O}(\alpha_s)+ {\cal O}(1/Q),\eqno(3.1.2)$$
${\cal O}(\alpha_s)$ and ${\cal O}(1/Q)$ are the QCD radiative
corrections and higher twist contributions. Inserting it to Eq.
(3.1.1), we obtain the relation between the spin structure functions
and the polarized quark distributions

$$g_1(x,Q^2)=\frac{1}{2}\sum_q e_q^2[\delta q(x,Q^2)+\delta \overline{q}(x,Q^2)]+{\cal O}(\alpha_s)+{\cal O}(1/Q). \eqno(3.1.3)$$
According to the renormalization group theory,

$$\frac{d g_1(x,Q^2)}{d\ln \mu_F}=0,\eqno(3.1.4)$$ it gives the DGLAP equation

$$Q^2\frac{d}{dQ^2}\delta q(x,Q^2)=\int^1_0\frac{dy}{y}\sum_{q'}\Delta P_{qq'}(x/y,\alpha_s(Q^2))\delta q'(y,Q^2),\eqno(3.1.5)$$
$\Delta P_{qq'}$ denotes the splitting functions. If we consider
only the leading order  $(LO$) approximation, we have

$$g_1^{DGLAP}(x,Q^2)=\frac{1}{2}\sum_q e_q^2[\delta q(x,Q^2)+\delta \overline{q}(x,Q^2)], \eqno(3.1.6)$$
These results are available at $Q^2>$ a few $GeV^2$.

    At lower $Q^2$, the multi-parton correlations are important and the inclusive lepton-nucleon cross section is dominated by complicate
higher twist terms. In fact, according to the operator product
expansion (OPE), the spin structure function in the proton
$g_1^p(x,Q^2)$ can be expressed as a series in $1/Q^2$,

$$g_1^p(x,Q^2)=g_1^{LT}(x,Q^2)+g_1^{HT}(x,Q^2).  \eqno(3.1.7)$$ The leading (twist-2) term corresponds to scattering
from a single free parton, while higher twist terms correspond to
multi-parton interactions. Only a little of higher twist can been
calculated perturbatively in terms of quark and gluon degrees of
freedom. For example, the contributions of parton recombination at
initial (or finite) state to the DGLAP evolution equation have been
calculated at leading order [10,11,12,13] and we denote this result as
$g_1^{DGLAP+ZRS}(x,Q^2)$. However, we can neither perform nor
interpret a partonic calculation of the higher twist effects
containing the correlations between the initial and finite partons
since they break the factorization schema. In a certain kinematic
regime, some of such higher twist contributions to $g_1^p(x,Q^2)$
appear as observable hadronic phenomenon. In this case, we may chose
a suitable phenomenological model, even do a parametrization to
describe the corresponding higher twist effects.

    We try use the well known Vector Meson Dominance (VMD) model to
mimic the above mentioned higher twist corrections. The reasons are
as follows. The handbag diagram Fig.13a is a typical time ordered
diagram describing Eq. (3.1.1), where the quark propagators
connect with the probe and the target has only the forward
components and these propagators can be broken as shown in Eq. (3.1.1)
since they are on-mass-shell. The corresponding backward quark
propagators construct the cat's ear diagram Fig.13b, which are
neglected since these backward propagators are absorbed by the
target in the collinear approximation [38,39]. However, the
contributions of Fig. 13c can not been neglected at low $Q^2$ due to
the corrections of quark-antiquark pair, which interacts with the
target as a virtual vector meson if the transverse momentum $k_\perp
\sim Q$ of quark pair is not large and confinement effects are
essential. The contribution of Fig. 13c can not factorized as eq.
(3.1.1). We use a phenomenological VMD model [40,41,42] to "isolate" this
contribution from Fig. 13c. Traditionally, such VMD model was used
to explain the structure function at low $Q^2$ region [43,44]. We
denote this contribution as $g_1^{VMD}(x,Q^2)$.

     The more complicated corrections to $g_1$ at low $Q^2$ are from
the higher order QCD effects $\cal{O}(\alpha_s)$ and higher order
recombination. In principle, we need to consider $all$ these
contributions, while it's beyond our ability. Our motivation is that
if one finds empirically that higher order corrections are deduced with a suitable scale down to
low $Q^2(\sim \mu^2)$, then one can extend our
leading order analysis of structure function data to $\mu^2$. If,
the results are incompatible with the data, then the data can be
used to extract the higher order contributions.

    In consequence, at lower $Q^2$ we have

$$g_1(x,Q^2)\simeq Pg_1^{DGLAP+ZRS}(x,Q^2)+g_1^{VMD}(x,Q^2)+g_1^{HT}(x,Q^2),\eqno(3.1.8)$$
where $P$ is the probability of inelastic events via bare
photon-parton interaction, the last term is the remaining higher
twist corrections and we will neglect it at small $x$. Equation
(3.1.8) implies that although the polarized partons share all
nucleon's spin, the higher twist effects mix with the contributions
of partons in the measuring spin structure function at low $Q^2$. We
emphasize that $g_1^{VMD}(x,Q^2)$ and $g_1^{HT}(x,Q^2)$ are
irrelevant to the definition of the parton distributions because
they violate the factorization schema, therefore, their
contributions to $g_1(x,Q^2)$ will not change the discussions about
spin in our previous section, which are the results of the polarized
parton distributions in the proton.

\newpage
\begin{center}
\subsection{Contributions of parton distributions and VMD part}
\end{center}

    The contributions of the
polarized parton distributions of the proton to the spin structure
functions at low $Q^2$ are

$$g_1^{DGLAP+ZRS}(x,Q^2)=\frac{1}{2}\sum_q e_q^2[\delta q(x,Q^2)+\delta
\overline{q}(x,Q^2)].\eqno(3.2.1)$$ We assume that all parton
distributions are freezed at scale $Q^2$ if $Q^2\le \mu^2$. Based on
this assumption we avoid the un-physical singularities at $Q\sim
\Lambda_{QCD}$.

    We present $x$-dependence of $g_1^{DGLAP+ZRS}(x,Q^2)$ at
several values of $Q^2$ in Fig. 14. One can find the dramatic change
of the spin structure function at $x<10^{-3}$ from a flat form to
dramatically decreasing. Considering Fig. 3, we conclude that the
large gluon helicity effect leads to this phenomenon.

    As we have mentioned that the contribution
from the vector meson in virtual photon to $g_1^p$ at $Q^2<1GeV^2$
is necessary. According to the VMD model [43,44],

    $$xg_1^{VMD}(x,Q^2)=\frac{1}{8\pi}\frac{m^4_\rho Q^2}{\gamma_\rho^2(Q^2+m^2_v)^2}\Delta\sigma_{\rho p}(s),\eqno(3.2.2)$$
where $\gamma_\rho$ is the coupling constant of $\rho$ vector meson
and proton; We consider the contributions of $\rho$ meson since
$\gamma_\rho\ll\gamma_\omega\ll \gamma_\phi$; $x$ is a variable
defined as $x=Q^2/(s+Q^2-m^2_p)$ rather than a momentum fraction of
parton, s is the CMS energy square of the $\gamma p$ collision. The
cross-sections $\Delta \sigma_{\rho p}(s)$ is the total cross
section for the scattering of polarized meson with the nucleon,
unfortunately, they are unknown. Usually, the following Regge theory
[45] is used,

$$\Delta\sigma_{\rho p}(s)\sim s^{\lambda-1}, at~s
\rightarrow \infty. \eqno(3.2.3)$$ The extrapolation of $g_1^p$ from
the measured region down to $x\sim 0$ suggests us to assume that
$\lambda=1-\epsilon$ and $\epsilon\sim 0$ is a small positive
parameter due to the requirement of integrability of $g_1^p$ at
$x\rightarrow 0$. In this work, we take $\epsilon=0$. Thus, we have

$$g_1^{VMD}(x,Q^2)\simeq B\frac{m^2_\rho Q^2}{(Q^2+m^2_\rho)^2}x^{-1}(1-x)^7,\eqno(3.2.4)$$
where $B=0.03$ and the factor $(1-x)^7$ is due to the spectator
counting rules at high $x$ [46], and it restricts the application of
the VMD model in  small $x$ range.

    We read $m_\rho^4/(Q^2+m_\rho^2)^2$ in Eq. (3.2.4) as the probability of the
VMD event, therefore,

$$P=1-\frac {m_\rho^4}{(Q^2+m_\rho^2)^2}, \eqno(3.2.5)$$ in Eq.(3.1.8).

\newpage
\begin{center}
\subsection{Predictions for spin structure function $g_1^p$ at small
$x$}
\end{center}

     What is the asymptotic behavior of $g_1^p$? This is a broadly discussed
subject. We plot
$g_1^p(x,Q^2)=Pg_1^{DGLAP+ZRS}(x,Q^2)+g_1^{VMD}(x,Q^2)$ with
different values of $Q^2$ in Fig. 15. There are two different
asymptotic behaviors of $g_1^p$ at small $x$: the VMD behavior $\sim
x^{-1}$ at $Q^2<1 GeV^2$ and the large gluon helicity effect at
$Q^2>3 GeV^2$. Besides, $g_1^p$ presents the twist form of the two
asymptomatic behaviors above, which is the mixing result of the
nonperturbative and perturbative dynamics.

    We compare our predicted $g_1^p$ at $x>10^{-3}$ with the data [47] in
Fig. 16. These data on 2010 are more precise than the previous data.
Note that the values of $Q^2$ of every measured point are different
and they are taken from Table I of [47]. The theoretical curve is a
smooth connection among these points. This figure shows that the
pQCD evolution almost control the behavior of $g_1^p$ at
$x>10^{-3}$.

    On the other hand, the combination of nonperturbative and
perturbative dynamics at $x<10^{-3}$ leads to a dramatic change of
$g_1^p$ around $Q^2=1\sim 3 GeV^2$.  Unfortunately, there are only
several data with large uncertainty about $g_1^p$ in this range. In
Figs. 17 and 18 we collect the HERA early data [48,49] at $Q^2=1$,
$10 GeV^2$ which are un-generally used and compare them with our
predicted $g_1^p$. Figure 19 shows some of these data (trigon) and
the comparisons with our results (dark points). Figure 20 is the
$Q^2$-dependence of $g_1^p$ with fixed $x$, the data are taken from
[50]. One can find that our predicted $g_1^p$ are compatible with
these data, although more precise measurements are necessary.

     Finally, we compare our results with the new COMAPSS (primary)
data [14,15,16,17,18] at $Q^2<1GeV^2$, which show that $g_1^p$ presents a flat
asymptomatic form at $x<10^{-3}$. This seems to contradict with the
predicted strong rise of $g_1^p$ at $Q^2<1GeV^2$ in Fig.3.
However, in the COMPASS fixed target experiments there is a strong 
correlation between $x$ and $Q^2$, which makes it possible that low $x$ measurements are along with 
low $Q^2$. In Fig. 21, we take the average values of $Q^2$ for each probing values of $x$
(see Fig.1 in Ref.[15-18]). The results
are acceptable. Obviously, the measurements at different $x$ with
different values of $Q^2$ in the fixed target experiments mix two
different asymptomatic behaviors of $g_1^p$.

    We predict the stronger $Q^2$- and $x$-dependence of $g_1^p$ at $0.01<Q^2<3 GeV^2$
and $x<0.1$ due to the mixture of nonperturbative vector meson
interactions and the QCD evolution of the parton distributions in
Fig.15. For testing this prediction, the measurements of $g_1^p$
with fixed $x$ or $Q^2$ at low $Q^2$ are necessary. The planning
Electron-Ion Collider (EIC), for example, eRHIC [51] and EIC@HIAF
[52] can probe a broad low $Q^2<1GeV^2$-range, where we can check
the predicted behavior of $g_1^p$ at fixed $x$ or $Q^2$.

\newpage
\begin{center}
\subsection{Discussions}
\end{center}

    In general consideration, both the logs of $1/x$ and
$Q^2$ are equally important at small $x$ and low $Q^2$, and one
should sum the double logarithmic (DL) terms
$(\alpha_s\ln^2(1/x))^n$, which predict the singular behavior
$g_1^p\sim x^{-\lambda}$ $(\lambda>0)$. It means that the BFKL
equation [53,54,55,56,57,58] and its nonlinear corrections- the Balitsky-kovchegov
equation [59,60,61] and the JIMWLK equation [62,63,64,65,66,67] should combine with
the DGLAP equation. However, the translation between the BFKL
equation and the DGLAP equation is a complicated technic. One of such
method is the Ciafaloni-Catani-Fiorani-Marchesini (CCFM) equation
[68,69,70,71], which is derived based on the two-scale unintegrated gluon
distribution. The solution of the CCFM equation is much more
complicated and has only proven to be practical with Monte Carlo
generators. To avoid this difficulty, some special methods
are proposed [72,73]. For example, the double logarithmic terms are
taken into account via a suitable kernel of the evolution equations
in the infrared evolution equations, which was first suggested by
Lipatov [74,75], or alternatively taking a singular initial parton
distributions at $x<10^{-2}$, one can also mimic the results of the
DL-resummation.

     In this work, the behavior of $g_1^p$ at the same range is obtained
through a long evolution of the DGLAP equation with the parton
recombination corrections. We find that it is different from the
predictions of the DL-resummation, the asymptomatic behavior of the
polarized quark distributions at $x\rightarrow 0$ is controlled by
$\Delta P_{qg}$ in the DGLAP equation, rather than the
$\ln^k(1/x)$-corrections to the DGLAP-kernel. Thus, the difficult DL
resummation can be replaced by the fits of the initial quark
distributions $\delta q_v(x,\mu^2)$ in the DGLAP equation if the
evolution distance is long enough. This conclusion was also obtained
in the unpolarized structure functions [76].

\newpage
\begin{center}
\section {Origins of the generalized Gerasimov-Drell-Hearn sum rule}

\subsection{Spin structure functions in the full $Q^2$ range}
\end{center}

    The Gerasimov-Drell-Hearn (GDH) sum rule reads

$$I_1(0)=\lim_{Q^2\rightarrow 0}\frac{2M^2}{Q^2}\Gamma_1(Q^2)=
-\frac{\kappa^2}{4}\sim -0.8,\eqno(4.1.1)$$where $\kappa$ is the
anomalous magnetic moment of the nucleon. On the other hand, the
Bjorken sum rule [77] says

$$\lim_{Q^2\rightarrow \infty}[\Gamma_1^p(Q^2)-\Gamma_1^n(Q^2)]=\frac{1}{6}\left\vert\frac{g_A}{g_V}\right\vert,\eqno(4.1.2)$$
this ratio is  accurately known as[78]:
$g_A/g_V=-1.2695\pm0.0029$.

The connection of the two sum rules by means of the generalized GDH
sum rule is(For an overview, see Ref. [79] for example)

$$I_1(Q^2)=\frac{2M^2}{Q^2}\Gamma_1(Q^2),\eqno(4.1.3)$$which
allows us to study the transition between the perturbative partonic
structure and nonperturbative hadronic picture of nucleon in
lepton-nucleon scattering processes. The data show that this sum
rule at low $Q^2<1GeV^2$ changes dramatically and exceeds the
variation bound at higher $Q^2$, which has been parameterized (but
not explanation) in [80,81,82,83]. The explanation of the generalized GDH sum
rule is an active subject.  For example, the phenomenological
constituent quark model [84,85], the VMD model [86,87], the resonance
contributions [88], the chiral perturbation theory ($\chi$PT) [89,90,91]
are used to understand the generalized GDH sum rule.

      The first moment of $g_1^p$ is

$$\Gamma_1^p(Q^2)=\Gamma_1^{DGLAP+ZRS}(Q^2)+\Gamma_1^{VMD}(Q^2)+\Gamma_1^{HT}(Q^2).\eqno(4.1.4)$$
From Eqs. (3.1.8) and (3.2.5) we obtain

$$\Gamma^{DGLAP+ZRS}(Q^2)\simeq 0.123\left(1-\frac{m^4_v}{(Q^2+m^2_v)^2}\right).\eqno(4.1.5)$$

    The dashed curve in Fig. 22 is our predicted
$\Gamma_1^{DGLAP+ZRS}(Q^2)$. On the other hand, we have

$$\Gamma_1^{VMD}(Q^2)=\int^1_0dxg_1^{VMD}(x,Q^2)\simeq
0.055\frac{m^2_vQ^2}{(Q^2+m^2_v)^2}. \eqno(4.1.6)$$

    Comparing the solid curve with $\Gamma_1^p(Q^2)$ data [92,93,94,95,96,97,98] in Fig. 22,
one can expect that the remaining higher twist corrections
$g_1^{HT}$ play a significant role at low $Q^2$ to the general GDH
sum rule. We will discuss them in detail next section.

\newpage
\begin{center}
\subsection{Higher twist contributions to the GDH sum rule}
\end{center}

    According to the OPE, the
appearance of scaling violations at low $Q^2$ is related to the
higher twist corrections to moments of structure functions. Higher
twists are expressed as matrix elements of operators involving
nonperturbative interactions between quarks and gluons. The study of
higher twist corrections gives us a direct insight into the nature
of long-range quark-gluon correlations. The higher twist corrections
to $g_1$ have several representations. In this work, we will try to
expose the remaining power suppression corrections to $\Gamma_1^p$.
For this sake, we make $\Gamma_1^{HT}(Q^2)$ (i.e., the data points
in Fig. 22)-[$\Gamma_1^{DGLAP+ZRS}(Q^2) +\Gamma_1^{VMD}(Q^2)$].
Figure 23 shows such a result at $Q^2>0.2~GeV^2$, which has been
smoothed with minimum $\chi^2/D.o.f.$.

    To expose the possible physical information of the curve in Fig. 23, according to QCD
operator product $1/Q^2$-expansion,

$$\Gamma_1^{HT}(Q^2)=\sum_{i=2}^{\infty}\frac{\mu_{2i}(Q^2)}{Q^{2i-2}},\eqno(4.2.1)$$ we take first three approximations

$$\Gamma_1^{HT(4)}(Q^2)=\frac{\mu_4(Q^2)}{Q^2},\eqno(4.2.2)$$

$$\Gamma_1^{HT(4+6)}(Q^2)=\frac{\mu_6(Q^2)+\mu_4(Q^2)Q^2}{Q^4},\eqno(4.2.3)$$

$$\Gamma_1^{HT(4+6+8)}(Q^2)=\frac{\mu_8(Q^2)+\mu_6(Q^2)Q^2+\mu_4(Q^2)Q^4}{Q^6}.\eqno(4.2.4)$$

    Then we plot the curves $Q^2\Gamma_1^{HT(4)}(Q^2)$, $Q^4\Gamma_1^{HT(4+6)}(Q^2)$ and  $Q^6\Gamma_1^{HT(4+6+8)}(Q^2)$
in Fig. 24. There are following interesting properties of these
results:

 (i) $Q^6\Gamma_1^{HT(4+6+8)}(Q^2)\rightarrow 0$, if $Q^2\rightarrow 0$. This implies that $\mu_8$ vanishes if it
is independent of $Q^2$. Therefore, $\Gamma_1^{HT(4+6)}(Q^2)$ is an
appropriate approximation.

 (ii) Three curves in Fig. 24 cross at a same point $Q^2\sim 1 GeV^2$.
Particularly, the intercept $\mu_6$ of the line suddenly changes its
value from -0.037 at $Q^2>1~GeV^2$ to 0.006 at $Q^2<1~GeV^2$. This
result exposes that the correlation among partons in the proton has
an obvious change near $Q\sim 1 GeV$.

 (iii)  We use

$$\Gamma_1^{HT(4+6)}(Q^2)=\frac{\mu_4}{Q^2+\epsilon^2}+\frac{\mu_6}{(Q^2+\epsilon^2)^2}~~at~Q^2<0.3 GeV^2\eqno(4.2.5)$$ to fit the data
at $Q^2<0.3 GeV^2$, where we add a parameter $\epsilon$ to remove
the unnatural singularity at $Q^2=0$. The value of $\epsilon$ is
sensitive to $I(0)$. We find that $\mu_4=-0.13GeV^2$,
$\mu_6=0.0528GeV^4 $ and $\epsilon^2=0.422 GeV^2$.

        In summary,

$$\Gamma_1^p(Q^2)=0.123\left(1-\frac{m^4_v}{(Q^2+m^2_v)^2}\right)+0.055\frac{m^2_vQ^2}{(Q^2+m^2_v)^2}+\Gamma_1^{HT}(Q^2)\eqno(4.2.6)$$where
the HT contributions are

$$\Gamma_1^{HT}(Q^2)=\left\{
\begin{array}{ll}
\frac{0.004M^2}{Q^2}-\frac{0.037M^4}{Q^4}~~at~Q^2>1 GeV^2\\
-\frac{0.048M^2}{Q^2}+\frac{0.0073M^4}{Q^4}~~at~0.3<Q^2<1 GeV^2\\
-\frac{0.13M^2}{Q^2+0.422M^2}+\frac{0.0528M^4}{(Q^2+0.422M^2)^2}~~at~Q^2<0.3~GeV^2
\end{array}
\right.,\eqno(4.2.7)$$ where $M^2=1GeV^2$. We present the comparison
of our $\Gamma_1^p(Q^2)$ with the data [92,93,94,95,96,97,98] in Fig. 25. The
corresponding $I_1^p(Q^2)$ is presented in Fig. 26.

\newpage
\begin{center}
\subsection{Discussions}
\end{center}

  The parton-hadron duality was first noted by Bloom and
Gilman [99,100] in deep inelastic scattering (DIS) and has been
confirmed by many measurements. At low energies (or intermediate
Bjorken variable $x$ and low $Q^2$) DIS reactions are characterized
by excitation of nucleon resonances; while at high virtuality such
processes have a partonic description. The smooth high-energy
scaling curve essentially reproduces the average of the resonance
peaks seen at low energies. Burkert and Ioffe [81] indicated that
the contribution of the isobar $\Delta(1232)$ electro-production at
small $Q^2$ can describe the general GDH sum rule, and they gave

$$\frac{\mu_4}{M^2}=-0.056\sim -0.063, at ~Q^2=0.3\sim 0.8 ~GeV^2\eqno(4.3.1)$$

$$\frac{\mu_6}{M^4}=0.010\sim 0.011, at~ Q^2=0.3\sim0.8 ~GeV^2, \eqno(4.3.2)$$ which
are compatible with our prediction Eq. (4.2.7).

    Our results in Fig.24 indicate that the negative twist-6
and twist-4 effects dominate the suppression of $\Gamma_1^p(Q^2)$ at
$Q^2>1 GeV^2$ and $Q^2<1GeV^2$. Particularly, the slope of $\mu_4$
of the lines, which cuts $Q^4\Gamma_1^{HT(4+6)}(Q^2)$ suddenly
changes its sign at $Q^2<1~GeV^2$. This result exposes that the
correlation among partons in the proton become stronger at scale
$\sim 0.2~fm$. We noted that Petronzio1, Simula and Ricco [101]
reported that the inelastic proton data obtained at Jefferson Lab
exhibit a possible extended objects with size of $\simeq 0.2-0.3~
fm$ inside the proton.

\newpage
\begin{center}
\section {Summary}
\end{center}

    In this work we consider that the nucleon is consisted of quarks
and gluons (partons) via QCD interactions even at low $Q^2$. A
general worry is that the correlations among partons may break the
definition of parton distributions and their evolution rules. As a
model, we treat these high twist effects as two parts: (i) The
leading recombination among initial partons, which modifies the
DGLAP equation but keeps the momentum conservation in Eqs.
(2.2.1)-(2.2.3) and the nucleon helicity sum rule Eq. (2.3.2); (ii)
The phenomenological VMD model and the parameterized higher twist
effects, which contribute to the measured structure functions of the
nucleon but they are irrelevant to the parton distributions in the
nucleon. In this framework, we discuss the electron scattering off a
nucleon at high energy in a special (Bjorken) infinite momentum
frame, where the virtual photon presents two components: bare photon
$\gamma^*$ and vector meson $V^*$with $J^{PG}=1^{--}$. In the former
case, $\gamma^*$ couples either with an on-mass-shell quark and
contributes $F_2^{DGLAP+ZRS}$, where we take a leading order
approximation and all higher order corrections are absorbed into the
free parameters, or with an off-mass-shell quark, which gives
$F_2^{HT}$. In the later case, the VMD model describes the
nonperturvative multi-parton interactions between $V^*$ and nucleon.
Thus, we present a compact theoretical model about the nucleon spin
structure.

    (i) We find that the gluon contribution to the spin of proton is much
larger than the predictions of most other theories. This result is
compatible with the recent NNPDF analysis and suggests a significant
orbital angular momentum of gluons to balance the contribution of
gluon spin. In concretely, the total proton spin at a bound state
scale $\mu^2$ is composed by $\sim 30\%$ quark spin and $\sim 70\%$
orbital angular momentum of the quarks, where two u-valence quarks
are rotating around a d-valence quark. With increasing $Q^2$, the
omitted gluons accumulate a larger positive helicity, which is
mainly balanced by their orbital momentum. Therefore, there are two
rotating groups in a polarized proton at $Q^2$: a slower quark group
and a faster gluon.

    (ii) We use the DGLAP equation with the parton recombination
corrections and the nonperturbative VMD model to predict the spin
structure functions $g_1^p$ of the proton. We first present a
complete picture for the translation of $g_1^p$ from low $Q^2(\sim
0)$ to high $Q^2$ at small $x$. We find that the contribution of the
large gluon helicity dominates $g_1^p$ at $x>10^{-3}$, but the
mixture with nonperturbative component complicates the asymptomatic
behavior of $g_1^p$ at $x<10^{-3}$. The results are compatible with
the data including the early HERA estimations and COMPASS new
results. The predicted strong $Q^2$- and $x$-dependence of $g_1^p$
at $0.01<Q^2<3 GeV^2$ and $x<0.1$ due to the mixture of
nonperturbative vector meson interactions and the QCD evolution of
the parton distributions can be checked on the next Electron-Ion
Collider (EIC).

    (iii) We discuss the contributions of parton distributions and
VMD component to the lowest moment of the spin-dependent proton
structure function. After removing the above two contributions from
the existing experimental data for $\Gamma_1^p(Q^2)$, the higher
twist power corrections present their interesting characters: parton
correlations at $Q^2\sim 1~GeV^2$ show a bend point, where the
twist-4 and twist-6 effects dominate the suppression of
$\Gamma_1^p(Q^2)$ at $Q^2<1 GeV^2$ and $Q^2>1 GeV^2$, respectively.
The results suggest a possible extended objects with size $0.2-0.3~
fm$ inside the proton. Within the analytic of these results, we are
able to achieve a rather good description of the data at all $Q^2$
region using a simple parameterized form of $\Gamma_1^p(Q^2)$.

\newpage
\noindent {\bf Appendix}:

    From Ref.[13], we have LO polarized gluon recombination functions

\begin{eqnarray*}
&& P_{g_+g_+\rightarrow g_+} =\frac{9}{4}{\frac { \left(
x_{{1}}+x_{{2}}-x \right) ^{3}}{x{x_{{2}}}^{ 2} \left(
x_{{1}}+x_{{2}} \right) ^{3}{x_{{1}}}^{2}} } ({x_{{1}}}^{
4}-2\,{x_{{1}}}^{3}x+{x_{{1}}}^{2}{x}^{2}+{x_{{2}}}^{4}-2\,{x_{{2}}}^{
3}x+{x_{{2}}}^{2}{x}^{2}\nonumber \\
&&\hspace{30mm}+{x_{{1}}}^{2}{x_{{2}}}^{2}-{x_{{1}}}^{2}x_{{2
}}x-x_{{1}}{x_{{2}}}^{2}x+x_{{1}}x_{{2}}{x}^{2} )\hspace{50mm}\rlap{(A.1)}\\
&&P_{ g_+g_+\rightarrow g_-} =\frac{9}{4}\frac { \left(
x_{{1}}+x_{{2}}-x \right)}{x{x_{{2}}}^{2} \left( x_{{1}}+x _{{2}}
\right) ^{3}{x_{{1}}}^{2}} (6\,{x_{{1}}}^{4
}x_{{2}}x+6\,{x_{{1}}}^{3}{x_{{2}}}^{2}x-3\,{x_{{1}}}^{3}x_{{2}}{x}^{2
}-7\,{x_{{1}}}^{2}x_{{2}}{x}^{3}\nonumber \\
&&\hspace{30mm}+11\,{x_{{1}}}^{2}{x_{{2}}}^{2}{x}^{2}
+6\,{x_{{2}}}^{4}x_{{1}}x+6\,{x_{{2}}}^{3}{x_{{1}}}^{2}x-3\,{x_{{2}}}^
{3}x_{{1}}{x}^{2}-7\,{x_{{2}}}^{2}x_{{1}}{x}^{3}+2\,x_{{1}}x_{{2}}{x}^
{4}\nonumber \\
&&\hspace{30mm}+{x_{{1}}}^{6}+{x_{{2}}}^{6}+2\,{x_{{1}}}^{5}x_{{2}}+2\,{x_{{1}}}^{
4}{x_{{2}}}^{2}-{x_{{1}}}^{4}{x}^{2}-2\,{x_{{1}}}^{3}{x}^{3}+2\,{x_{{1
}}}^{2}{x}^{4}+2\,{x_{{2}}}^{5}x_{{1}}\nonumber \\
&&\hspace{30mm}+2\,{x_{{2}}}^{4}{x_{{1}}}^{2}-{
x_{{2}}}^{4}{x}^{2}-2\,{x_{{2}}}^{3}{x}^{3}+2\,{x_{{2}}}^{2}{x}^{4}+2
\,{x_{{1}}}^{3}{x_{{2}}}^{3})\hspace{30mm}\rlap{(A.2)}
\\
&&P_{ g_+g_-\rightarrow g_+} ={\frac {9}{4}}\,{\frac
{(x_{{1}}+x_{{2}}-x)}{x{x_{{2}}}^ {2}\left (x_{{1}}+x_{{2}}\right
)^{7}{x_{{1}}}^{2}}}(141\,
{x_{{1}}}^{7}{x}^{2}x_{{2}}+42\,{x_{{1}}}^{4}{x}^{4}{x_{{2}}}^{2}-100
\,{x_{{1}}}^{8}xx_{{2}}\nonumber\\
 &&\hspace{30mm}+19\,{x_{{1}}}^{5}{x}^{4}x_{{2}}-5\,{x_{{2}}}^{
8}x_{{1}}x+39\,{x_{{1}}}^{2}{x}^{4}{x_{{2}}}^{4}+137\,{x_{{1}}}^{3}{x_
{{2}}}^{6}x-86\,{x_{{1}}}^{6}{x}^{3}x_{{2}}\nonumber\\
&&\hspace{30mm} +26\,{x_{{2}}}^{9}x_{{1}}+5
\,{x_{{1}}}^{10}+5\,{x_{{2}}}^{10}+155\,{x_{{1}}}^{4}{x}^{2}{x_{{2}}}^
{4}-40\,{x_{{1}}}^{2}{x_{{2}}}^{5}{x}^{3}-212\,{x_{{1}}}^{6}x{x_{{2}}}
^{3}\nonumber\\
&&\hspace{30mm}
-124\,{x_{{1}}}^{3}{x}^{3}{x_{{2}}}^{4}-196\,{x_{{1}}}^{4}{x}^{3}{
x_{{2}}}^{3}-79\,{x_{{1}}}^{2}{x_{{2}}}^{6}{x}^{2}+128\,{x_{{1}}}^{4}{
x_{{2}}}^{5}x+44\,{x_{{2}}}^{7}{x_{{1}}}^{2}x\nonumber\\
&&\hspace{30mm}-47\,{x_{{1}}}^{5}x{x_{{2
}}}^{4}-177\,{x_{{1}}}^{5}{x}^{3}{x_{{2}}}^{2}-209\,{x_{{1}}}^{7}x{x_{
{2}}}^{2}+40\,{x_{{1}}}^{3}{x}^{4}{x_{{2}}}^{3}-33\,{x_{{1}}}^{3}{x_{{
2}}}^{5}{x}^{2}\nonumber\\
&&\hspace{30mm}+{x_{{2}}}^{6}x_{{1}}{x}^{3}+319\,{x_{{1}}}^{5}{x}^{2}{
x_{{2}}}^{3}+291\,{x_{{1}}}^{6}{x}^{2}{x_{{2}}}^{2}-35\,{x_{{2}}}^{7}x
_{{1}}{x}^{2}+13\,{x_{{2}}}^{5}{x}^{4}x_{{1}}\nonumber\\
&&\hspace{30mm}+64\,{x_{{2}}}^{7}{x_{{1}
}}^{3}+57\,{x_{{2}}}^{8}{x_{{1}}}^{2}+26\,{x_{{1}}}^{9}x_{{2}}+57\,{x_
{{1}}}^{8}{x_{{2}}}^{2}+64\,{x_{{1}}}^{7}{x_{{2}}}^{3}+34\,{x_{{1}}}^{
6}{x_{{2}}}^{4}\nonumber\\
&&\hspace{30mm}+12\,{x_{{1}}}^{5}{x_{{2}}}^{5}+34\,{x_{{1}}}^{4}{x_{{2
}}}^{6}-4\,{x_{{2}}}^{9}x+2\,{x_{{2}}}^{6}{x}^{4}+2\,{x_{{2}}}^{7}{x}^
{3}-5\,{x_{{2}}}^{8}{x}^{2}+30\,{x_{{1}}}^{8}{x}^{2}\nonumber\\
&&\hspace{30mm}-20\,{x_{{1}}}^{7}
{x}^{3}+5\,{x_{{1}}}^{6}{x}^{4}-20\,{x_{{1}}}^{9}x )
\hspace{60mm}\rlap{(A.3)}
\\
&&P_{g_+g_-\rightarrow g_-}={\frac {9}{4}}\,{\frac {
(x_{{1}}+x_{{2}}-x )}{x{x_{{2}} }^{2} (x_{{1}}+x_{{2}}
)^{7}{x_{{1}}}^{2}}}(-31\,
{x_{{1}}}^{7}{x}^{2}x_{{2}}+27\,{x_{{1}}}^{4}{x}^{4}{x_{{2}}}^{2}-7\,{
x_{{1}}}^{8}xx_{{2}}+9\,{x_{{1}}}^{5}{x}^{4}x_{{2}}\nonumber\\
&&\hspace{30mm}-104\,{x_{{2}}}^{8}
x_{{1}}x+54\,{x_{{1}}}^{2}{x}^{4}{x_{{2}}}^{4}-206\,{x_{{1}}}^{3}{x_{{
2}}}^{6}x+3\,{x_{{1}}}^{6}{x}^{3}x_{{2}}+26\,{x_{{2}}}^{9}x_{{1}}+5\,{
x_{{1}}}^{10}\nonumber\\
&&\hspace{30mm}+5\,{x_{{2}}}^{10}+115\,{x_{{1}}}^{4}{x}^{2}{x_{{2}}}^{4}
-211\,{x_{{1}}}^{2}{x_{{2}}}^{5}{x}^{3}+125\,{x_{{1}}}^{6}x{x_{{2}}}^{
3}-192\,{x_{{1}}}^{3}{x}^{3}{x_{{2}}}^{4}\nonumber\\
&&\hspace{30mm}-92\,{x_{{1}}}^{4}{x}^{3}{x_{
{2}}}^{3}+319\,{x_{{1}}}^{2}{x_{{2}}}^{6}{x}^{2}-25\,{x_{{1}}}^{4}{x_{
{2}}}^{5}x-217\,{x_{{2}}}^{7}{x_{{1}}}^{2}x+136\,{x_{{1}}}^{5}x{x_{{2}
}}^{4}\nonumber\\
&&\hspace{30mm}-24\,{x_{{1}}}^{5}{x}^{3}{x_{{2}}}^{2}+34\,{x_{{1}}}^{7}x{x_{{2}
}}^{2}+36\,{x_{{1}}}^{3}{x}^{4}{x_{{2}}}^{3}+307\,{x_{{1}}}^{3}{x_{{2}
}}^{5}{x}^{2}-106\,{x_{{2}}}^{6}x_{{1}}{x}^{3}\nonumber\\
&&\hspace{30mm}-41\,{x_{{1}}}^{5}{x}^{2
}{x_{{2}}}^{3}-67\,{x_{{1}}}^{6}{x}^{2}{x_{{2}}}^{2}+157\,{x_{{2}}}^{7
}x_{{1}}{x}^{2}+27\,{x_{{2}}}^{5}{x}^{4}x_{{1}}+64\,{x_{{2}}}^{7}{x_{{
1}}}^{3}\nonumber\\
&&\hspace{30mm}+57\,{x_{{2}}}^{8}{x_{{1}}}^{2}+26\,{x_{{1}}}^{9}x_{{2}}
+57\,{
x_{{1}}}^{8}{x_{{2}}}^{2}+64\,{x_{{1}}}^{7}{x_{{2}}}^{3}+34\,{x_{{1}}}
^{6}{x_{{2}}}^{4}+12\,{x_{{1}}}^{5}{x_{{2}}}^{5}\nonumber\\
&&\hspace{30mm}+34\,{x_{{1}}}^{4}{x_{
{2}}}^{6}-20\,{x_{{2}}}^{9}x+5\,{x_{{2}}}^{6}{x}^{4}-20\,{x_{{2}}}^{7}
{x}^{3}+30\,{x_{{2}}}^{8}{x}^{2}-5\,{x_{{1}}}^{8}{x}^{2}+2\,{x_{{1}}}^
{7}{x}^{3}\nonumber\\
&&\hspace{30mm}+2\,{x_{{1}}}^{6}{x}^{4}-4\,{x_{{1}}}^{9}x)\hspace{82mm}\rlap{(A.4)}
\\
\end{eqnarray*}

    Setting $x_1=x_2=y$, one can find

$$\Delta P_{gg\rightarrow g}=[P_{g_+g_+\rightarrow g_+}-P_{g_+g_+\rightarrow
g_-}-P_{g_+g_-\rightarrow g_+}+P_{g_+g_-\rightarrow
g_-}]$$
$$=[P_{g_+g_+\rightarrow g_+}-P_{g_+g_+\rightarrow
g_-}+P_{g_+g_-\rightarrow g_+}-P_{g_+g_-\rightarrow
g_-}]$$
$$=\frac {27}{64}\frac {(2y-x)(-20y^3+12y^2x-x^3)}{y^5}$$

    Similarly,

\begin{eqnarray*}
&&P_{g_+g_+\rightarrow q_+} =\frac{1}{12}\frac { (x_{{1}}+x_{{2}}-x
)^{2}}{(x_{{1}}+x_{{2}} )^{3}{x_{{2}}}^{2}{x_{{1 }}}^{2}}
(4\,{x_{{1}}}
^{4}+7\,{x_{{1}}}^{3}x_{{2}}-8\,{x_{{1}}}^{3}x+2\,{x_{{1}}}^{2}{x_{{2}
}}^{2}-6\,{x_{{1}}}^{2}xx_{{2}}\nonumber\\
&&\hspace{30mm}+4\,{x}^{2}{x_{{1}}}^{2}+4\,{x}^{2}{x_{
{2}}}^{2}-x_{{1}}{x_{{2}}}^{3}+2\,x_{{1}}x{x_{{2}}}^{2}-x_{{1}}x_{{2}}
{x}^{2}) \hspace{28mm}\rlap{(A.5)}
\\
&&P_{g_+g_+\rightarrow q_-} =\frac{1}{12} \frac { (x_{{1}}+x_{{2}}-x
)^{2}}{(x_{{1}}+x_{{2}} )^{3}{x_{{2}}}^{2}{x_{{1}}}^{2}}
(4\,{x}^{2}{x
_{{1}}}^{2}+4\,{x_{{1}}}^{2}{x_{{2}}}^{2}+8\,x_{{1}}{x_{{2}}}^{3}-8\,x
_{{1}}x{x_{{2}}}^{2}+4\,{x_{{2}}}^{4}\nonumber\\
&&\hspace{30mm} -8\,{x_{{2}}}^{3}x+4\,{x}^{2}{x_{
{2}}}^{2}-x_{{1}}x_{{2}}{x}^{2}) \hspace{65mm}\rlap{(A.6)}
\\
&&P_{g_+g_-\rightarrow q_+} =\frac{1}{12}\frac { (x_{{1}}+x_{{2}}-x
)^{2}}{ (x _{{1}}+x_{{2}} )^{7}{x_{{2}}}^{2}{x_{{1}}}^{2}}
(4\,{x_{{1}}}
^{6}{x}^{2}+4\,{x_{{1}}}^{4}{x_{{2}}}^{4}-8\,{x_{{1}}}^{7}x+24\,{x_{{1
}}}^{6}{x_{{2}}}^{2}+16\,{x_{{1}}}^{5}{x_{{2}}}^{3}\nonumber\\
&&\hspace{30mm}+16\,{x_{{1}}}^{7}x
_{{2}}+4\,{x_{{1}}}^{8}+24\,{x}^{2}{x_{{2}}}^{5}x_{{1}}+4\,{x}^{2}{x_{
{2}}}^{6}-10\,{x_{{1}}}^{3}{x}^{2}{x_{{2}}}^{3}+33\,{x_{{1}}}^{2}{x}^{
2}{x_{{2}}}^{4}\nonumber\\
&&\hspace{30mm}-8\,{x_{{1}}}^{4}x{x_{{2}}}^{3}+24\,{x_{{1}}}^{5}{x}^{2
}x_{{2}}+33\,{x_{{1}}}^{4}{x}^{2}{x_{{2}}}^{2}+14\,{x_{{1}}}^{3}x{x_{{
2}}}^{4}-40\,{x_{{1}}}^{6}xx_{{2}}\nonumber\\
&&\hspace{30mm}-53\,{x_{{1}}}^{5}x{x_{{2}}}^{2} -8\,
x{x_{{2}}}^{5}{x_{{1}}}^{2}-9\,x_{{1}}x{x_{{2}}}^{6} )
\hspace{50mm}\rlap{(A.7)}
\\
&&P_{g_+g_-\rightarrow q_-} =\frac{1}{12}\frac { (x_{{1}}+x_{{2}}-x
)^{2}}{(x_{{1}}+x_{{2}} )^{7}{x_ {{2}}}^{2}{x_{{1}}}^{2}}
(24\,{x_{{2}}
}^{6}{x_{{1}}}^{2}+16\,{x_{{2}}}^{7}x_{{1}}-8\,{x_{{2}}}^{7}x+16\,{x_{
{2}}}^{5}{x_{{1}}}^{3}+4\,{x_{{2}}}^{8}\nonumber\\
&&\hspace{30mm}+4\,{x_{{1}}}^{6}{x}^{2}+4\,{x_
{{1}}}^{4}{x_{{2}}}^{4}+6\,{x}^{2}{x_{{2}}}^{5}x_{{1}}+4\,{x}^{2}{x_{{
2}}}^{6}+8\,{x_{{1}}}^{3}{x}^{2}{x_{{2}}}^{3}+15\,{x_{{1}}}^{2}{x}^{2}
{x_{{2}}}^{4}\nonumber\\
&&\hspace{30mm}-13\,{x_{{1}}}^{4}x{x_{{2}}}^{3}+24\,{x_{{1}}}^{5}{x}^{2}
x_{{2}}+51\,{x_{{1}}}^{4}{x}^{2}{x_{{2}}}^{2}-35\,{x_{{1}}}^{3}x{x_{{2
}}}^{4}+{x_{{1}}}^{5}x{x_{{2}}}^{2}\nonumber\\
&&\hspace{30mm}-35\,x{x_{{2}}}^{5}{x_{{1}}}^{2}-22
\,x_{{1}}x{x_{{2}}}^{6} ) \hspace{70mm}\rlap{(A.8)}
\\\end{eqnarray*}

    Setting $x_1=x_2=y$, we have

$$\Delta P_{gg\rightarrow q}=[P_{g_+g_+\rightarrow q_+}-P_{g_+g_+\rightarrow
q_-}-P_{g_+g_-\rightarrow q_+}+P_{g_+g_-\rightarrow
q_-}]$$
$$=[P_{g_+g_+\rightarrow q_+}-P_{g_+g_+\rightarrow
q_-}+P_{g_+g_-\rightarrow q_+}-P_{g_+g_-\rightarrow
q_-}]$$
$$=\frac {1}{48}\frac {(2y-x)^2
(-y+x)}{y^4}$$

\newpage
\begin{figure}[htp]
\centering
\vskip -3cm
\includegraphics[width=0.6\textwidth]{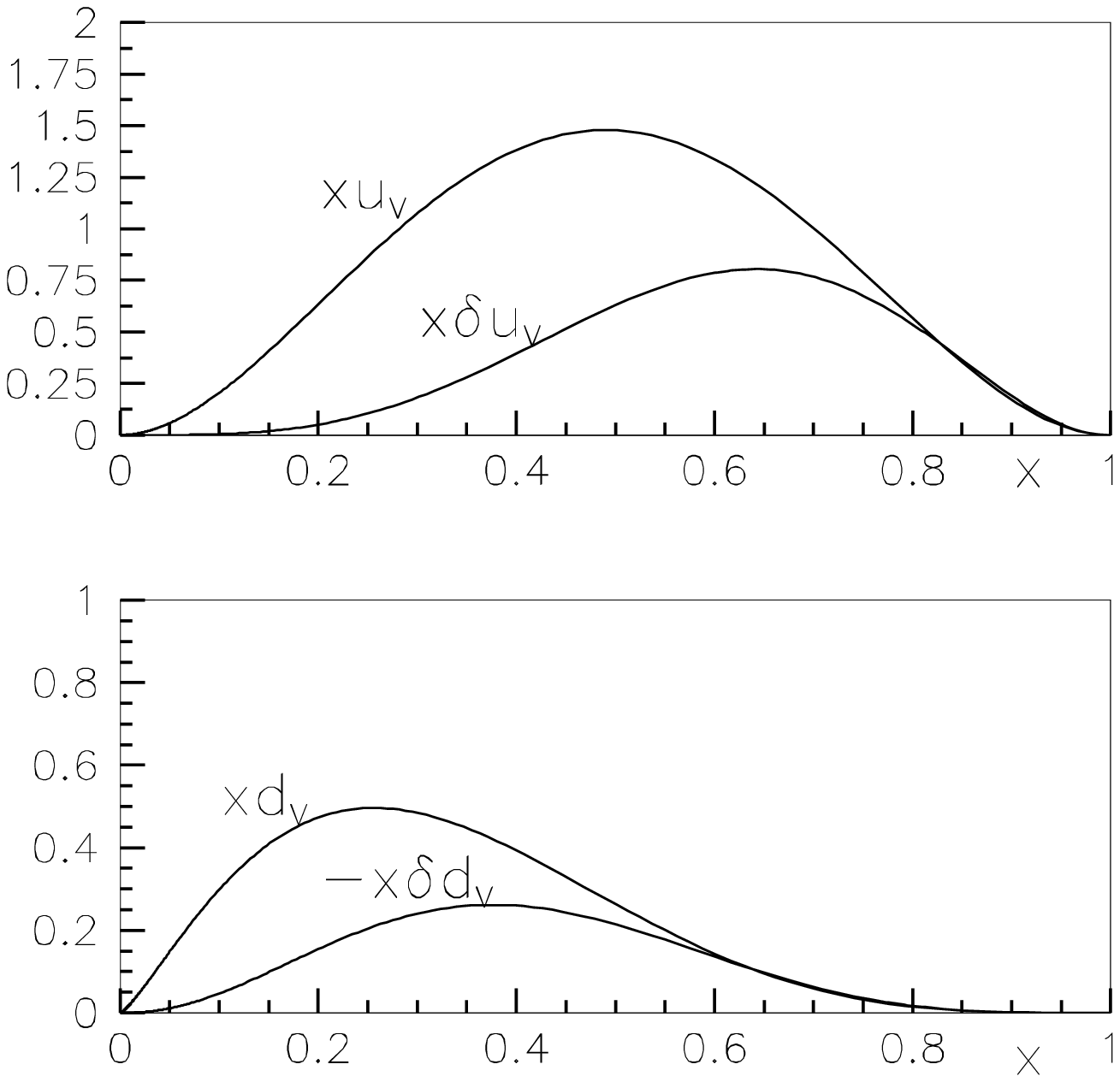}
\vskip -3cm
\caption{Input valence quark distributions at $\mu^2=0.064GeV^2$
for polarized and un-polarized densities.} \label{fig1}
\centering

\vskip -2cm
\includegraphics[width=0.6\textwidth]{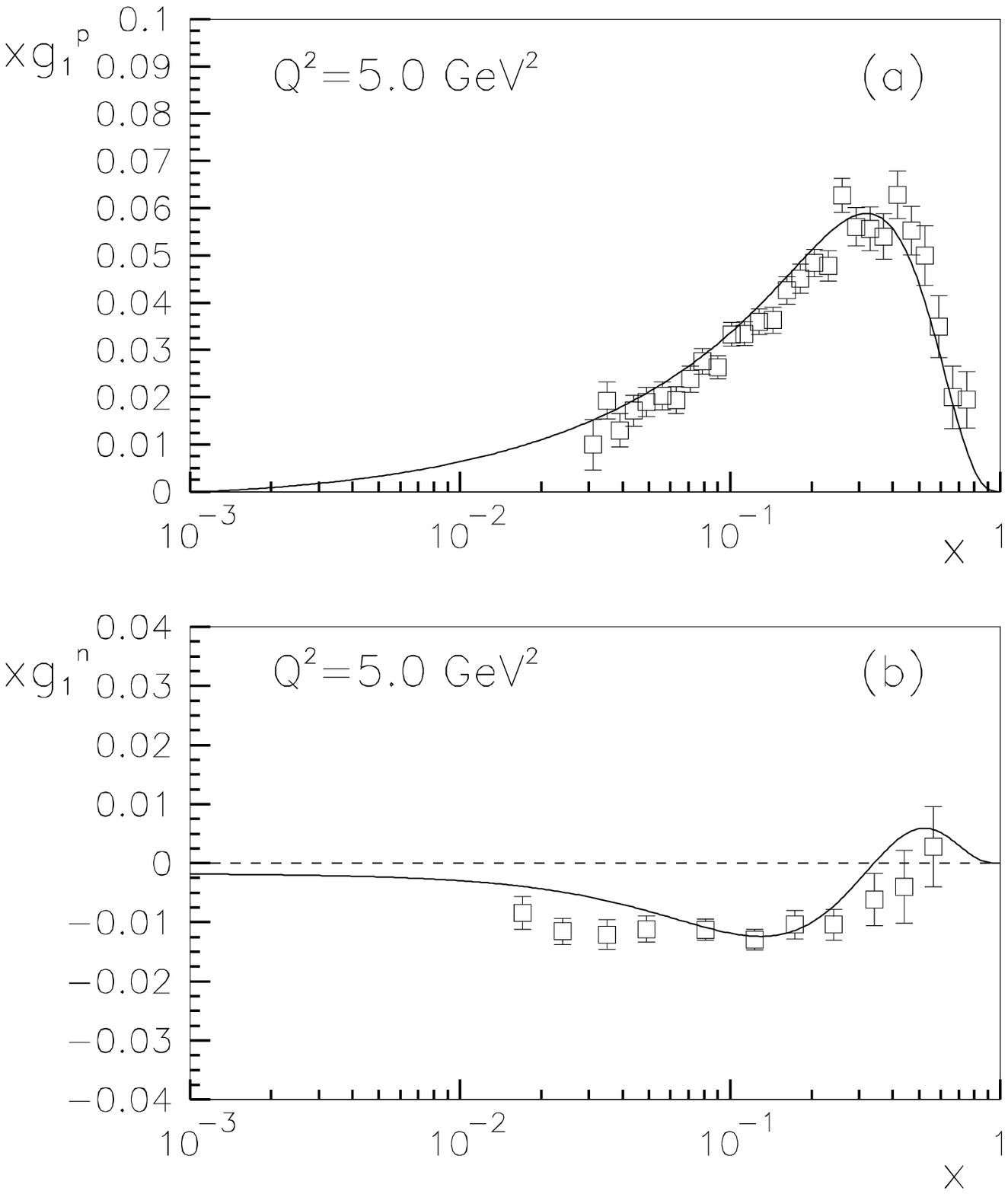}
\vskip -2cm
\caption{Fitting the data [21] for $g_1$ at $Q^2=5 GeV^2$ using the
input distributions Eqs. (2.2.1)-(2.2.6). } \label{fig2}
\end{figure}

\newpage

\begin{figure}[htp]
\centering
\vskip -3cm
\includegraphics[width=0.6\textwidth]{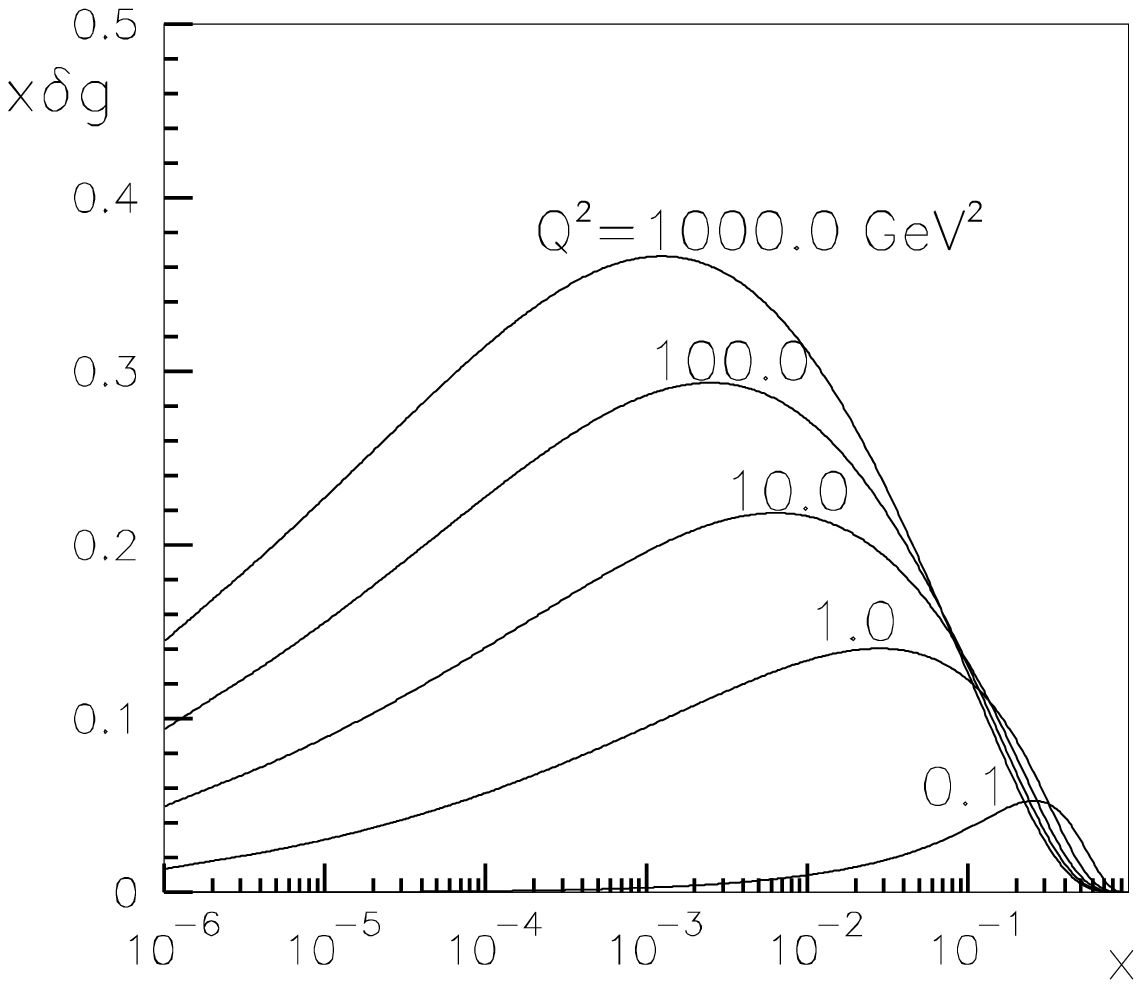}

\vskip -4cm
\caption{Predicted polarized gluon distribution $x\delta g(x,Q^2)$
in the proton at the different $Q^2$-scales, which show the
accumulation of radiative polarized gluons at small $x$ in the
evolution.} \label{fig3}

\centering

\vskip -1cm
\includegraphics[width=0.6\textwidth]{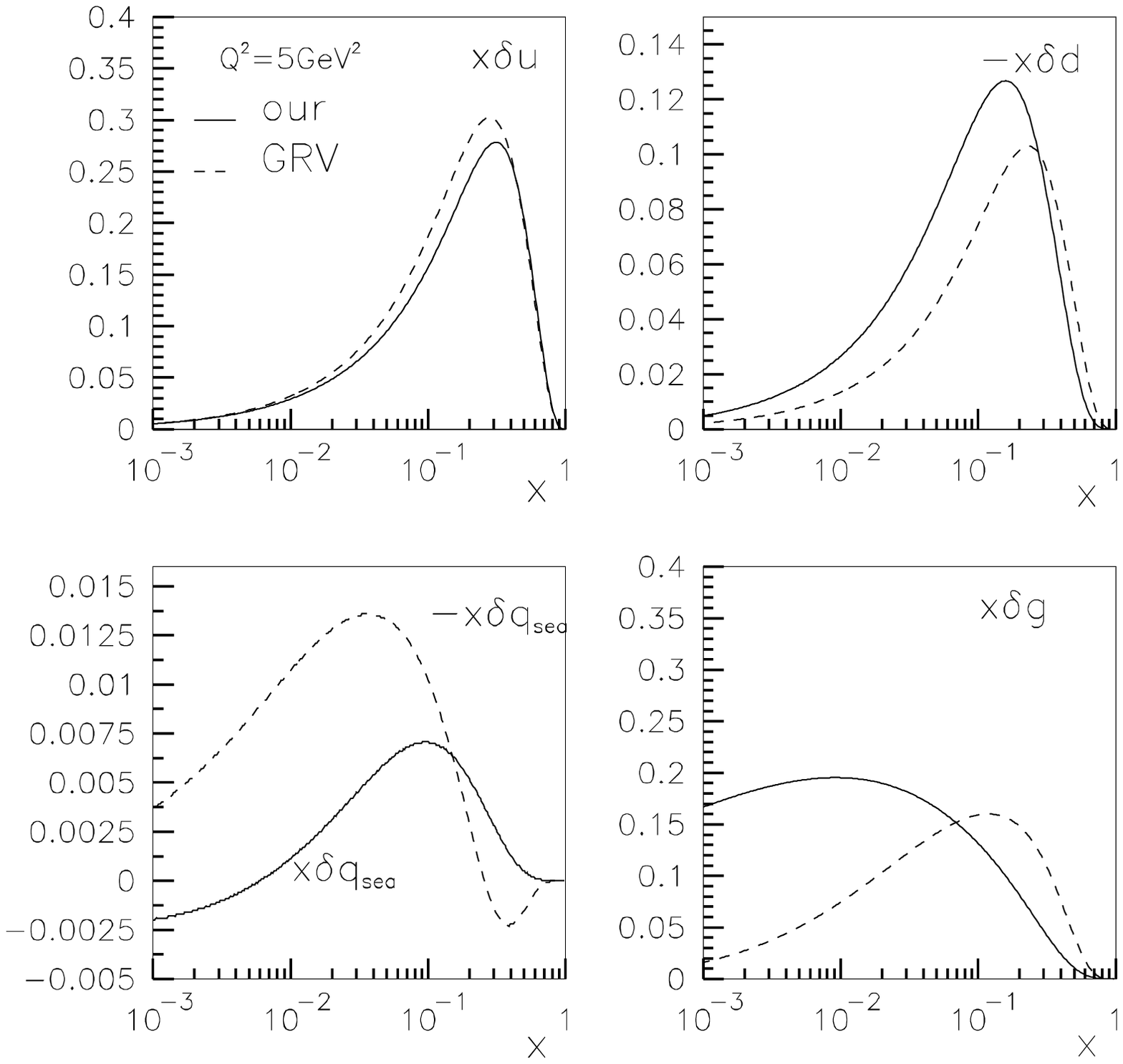}

\vskip -3cm
\caption{Comparisons of our predicted polarized LO parton
distributions at $Q^2=5GeV^2$ with the GRV distributions [22].}
\label{fig4}
\end{figure}

\newpage
\begin{figure}[htp]
\centering

\vskip -3cm
\includegraphics[width=0.6\textwidth]{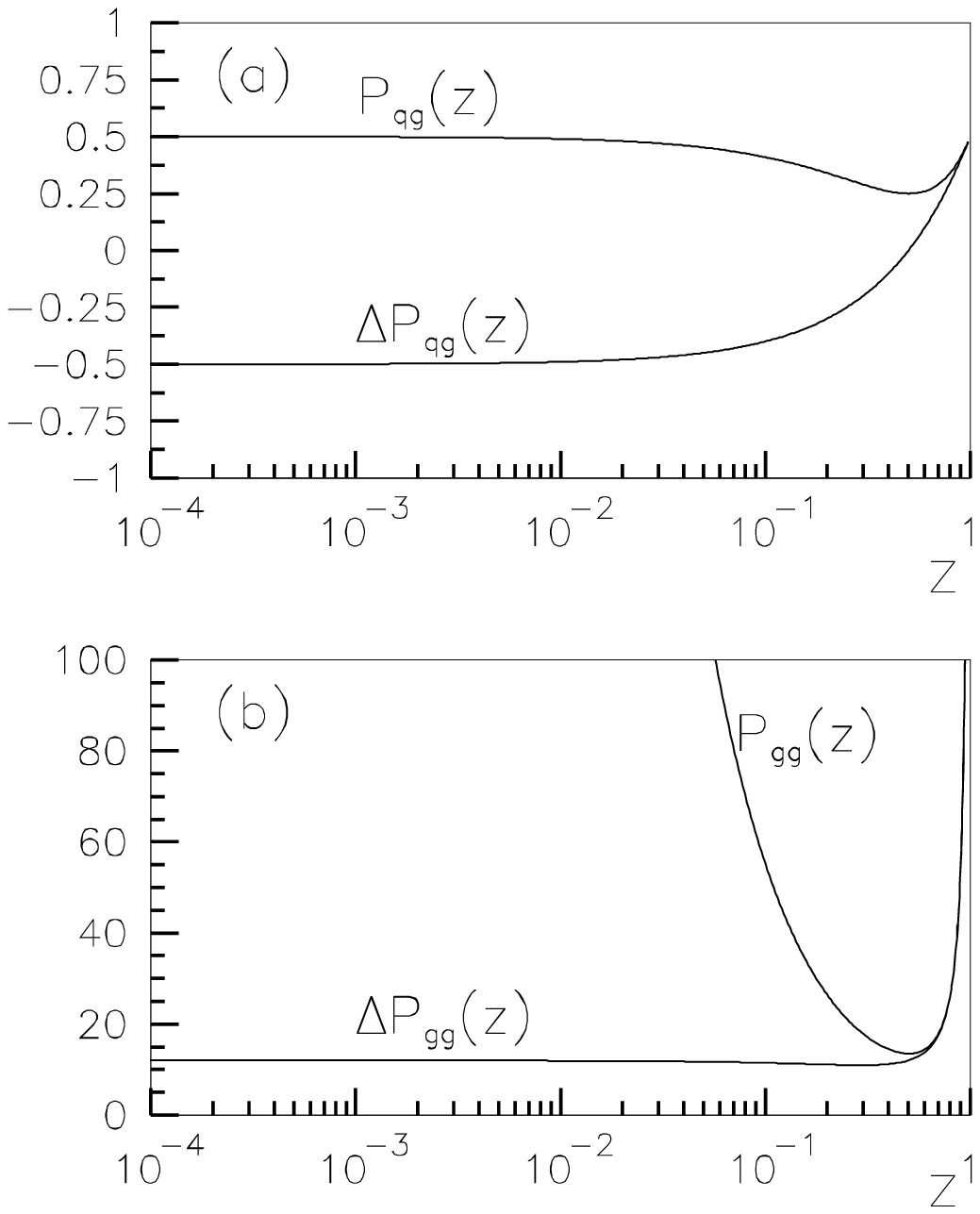}

\vskip -3cm
\caption{The splitting functions.}
\centering

\vskip -1cm
\includegraphics[width=0.6\textwidth]{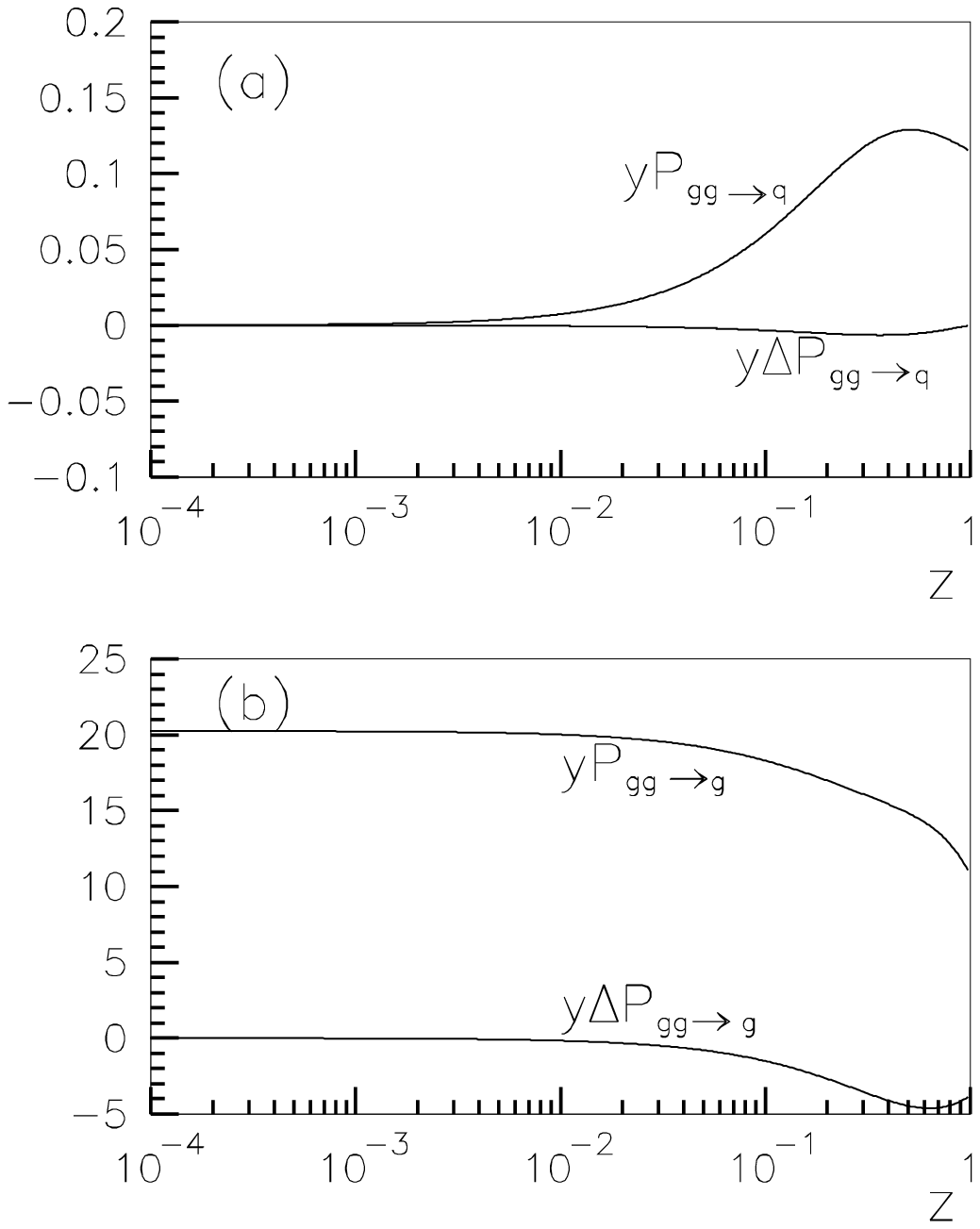}

\vskip -3cm
\caption{The recombination functions.} \label{fig6}
\end{figure}

\newpage
\begin{figure}[htp]
\centering

\vskip -3cm
\includegraphics[width=0.6\textwidth]{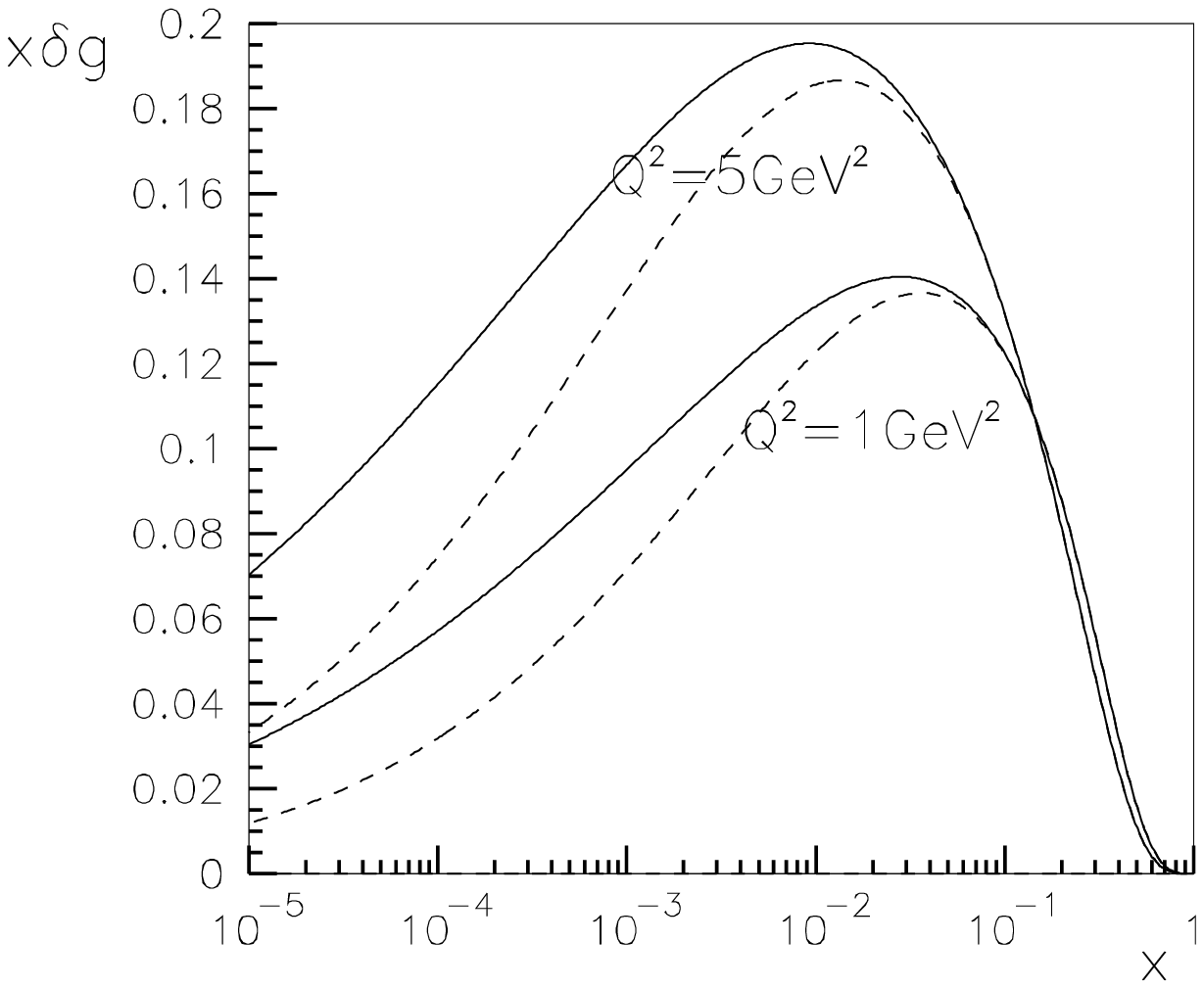}

\vskip -3cm
\caption{Predicted polarized gluon distributions $x\delta g$ in
the nucleon at $Q^2=1$ and $5 GeV^2$ with gluon recombination
corrections (solid curves) and without gluon recombination
corrections (dashed curves).} \label{fig7}

\centering

\vskip -2cm
\includegraphics[width=0.6\textwidth]{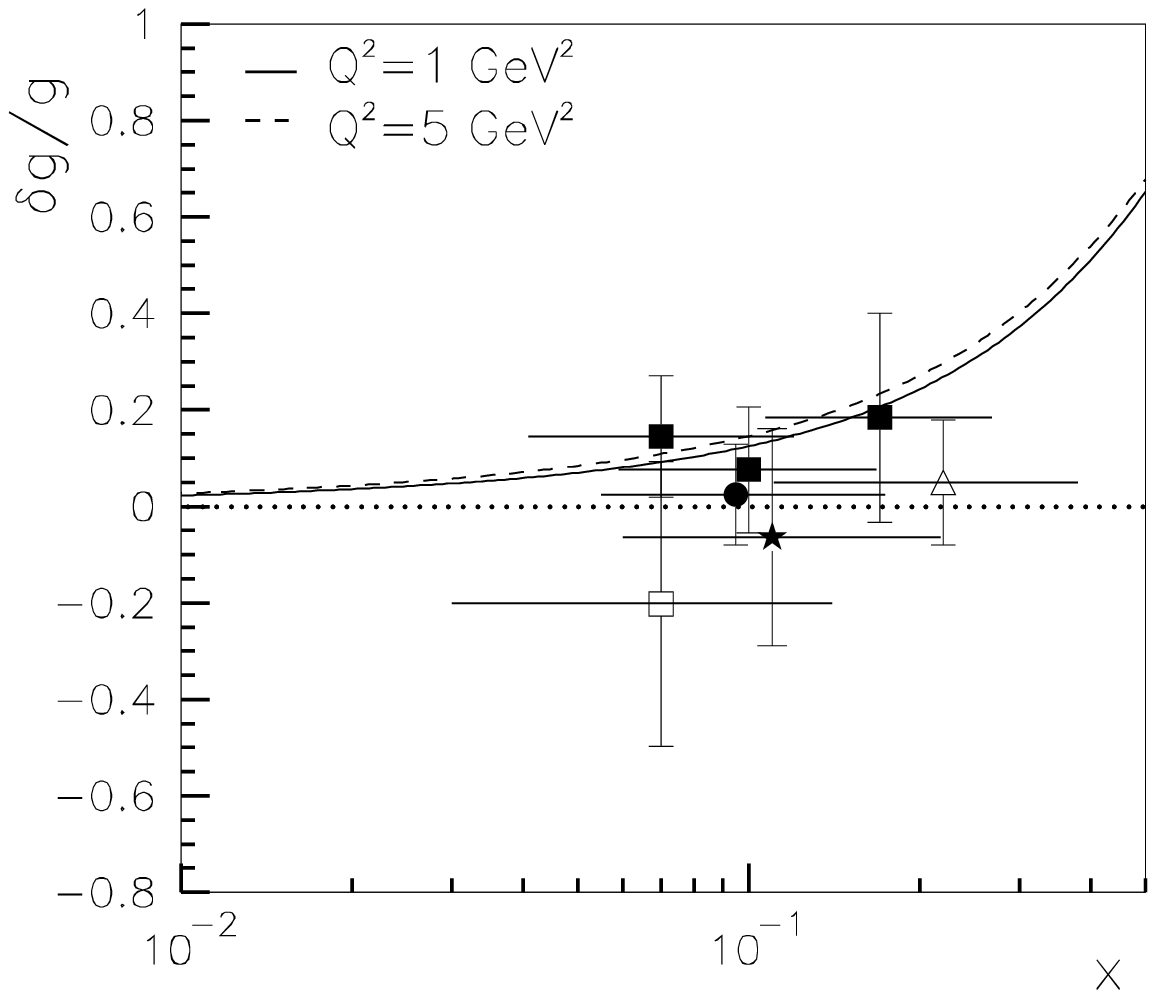}

\vskip -3cm
\caption{Comparison of dynamically predicted $\delta g/g$ with the
COMPASS data [23] at $Q^2=1$ and $5 GeV^2$. } \label{fig8}
\end{figure}

\newpage
\begin{figure}[htp]
\centering

\vskip -3cm
\includegraphics[width=0.6\textwidth]{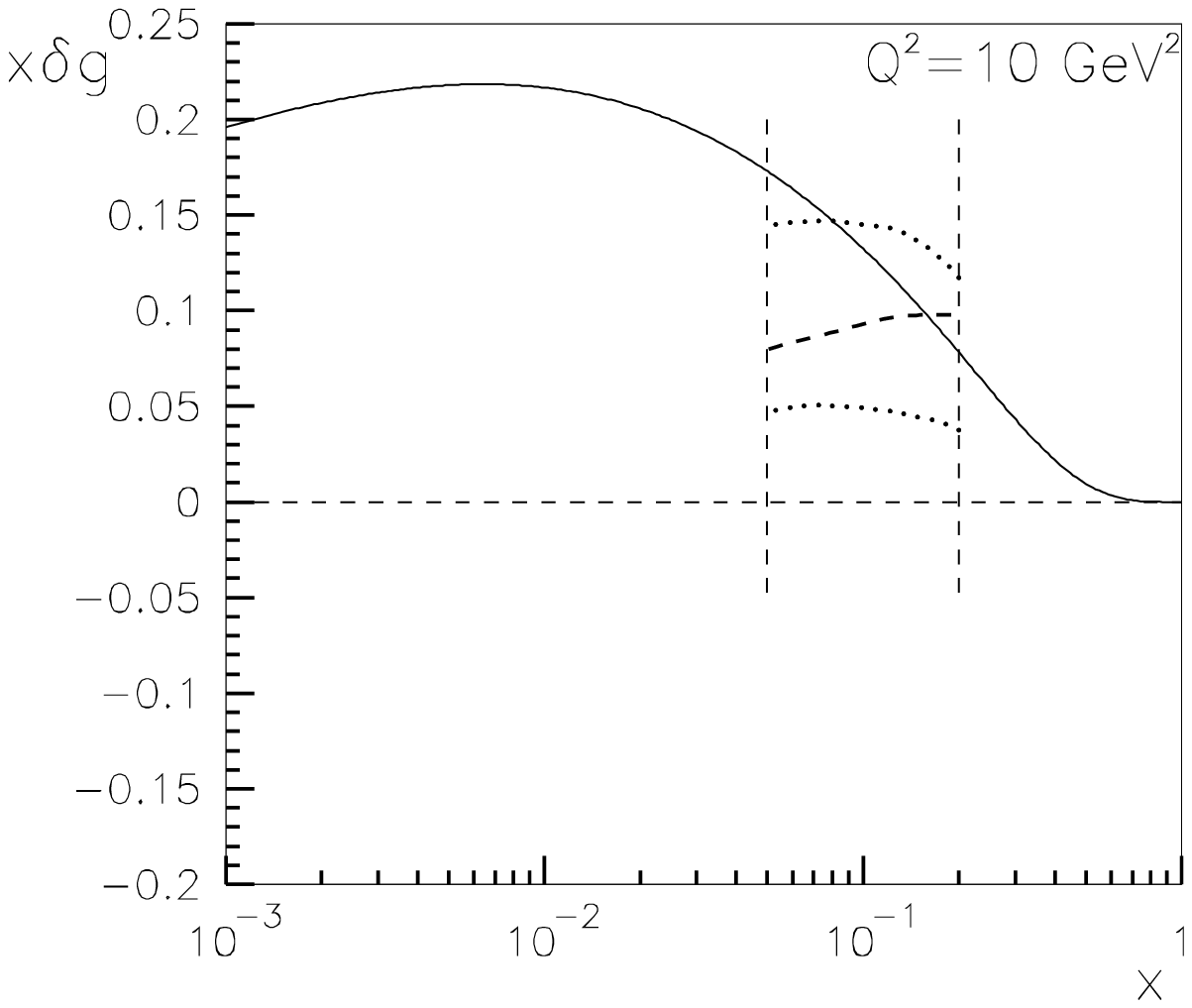}

\vskip -3cm
\caption{Our predicted $x\delta g$ at $Q^2=10 GeV^2$ (solid curve).
Broken curve is the result by DSSV using RHIC measurements in [24];
dotted curves are the fits within the $90\%$ confidence level (C.L.)
interval  } \label{fig9}

\centering

\vskip -2cm
\includegraphics[width=0.6\textwidth]{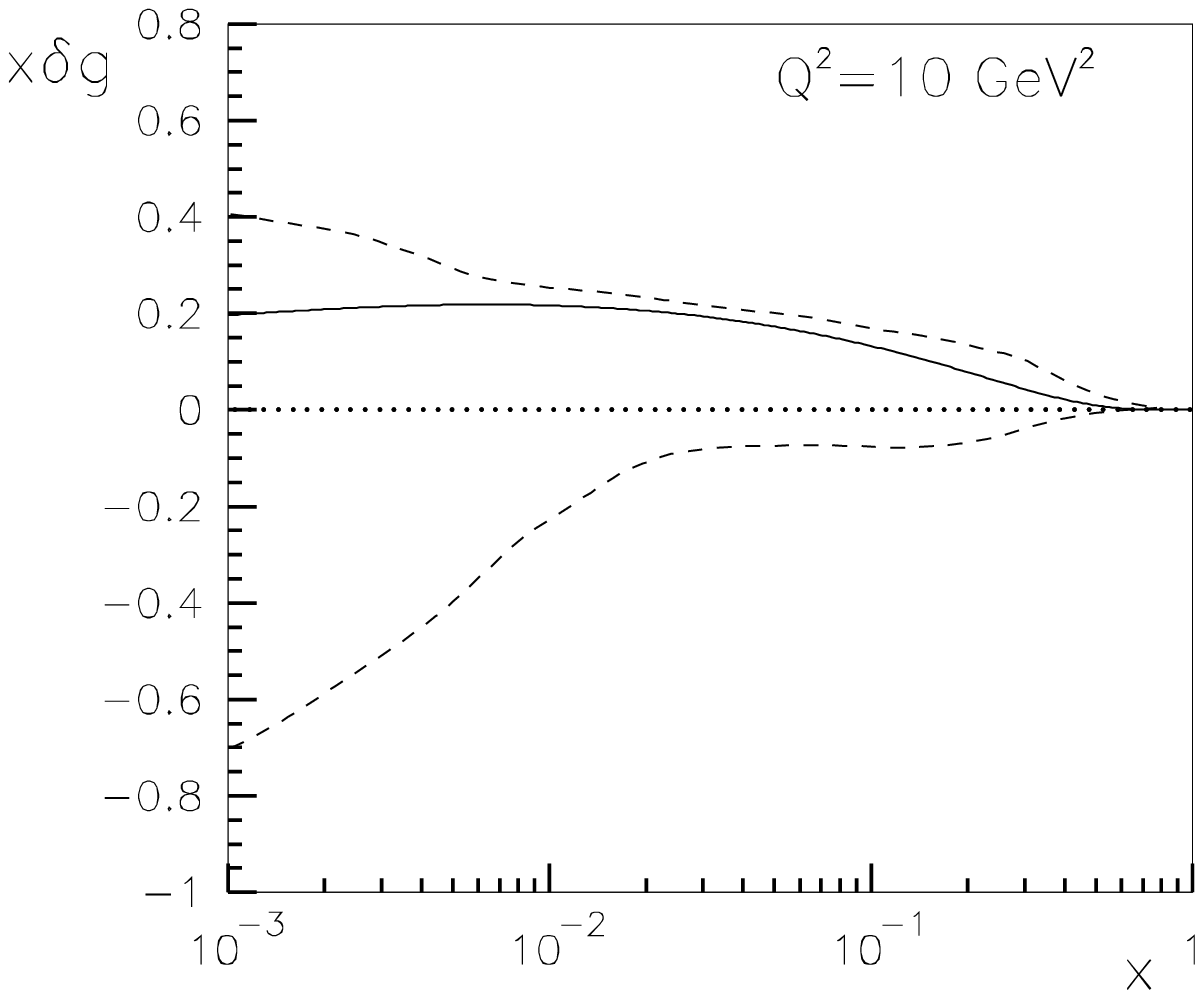}

\vskip -3cm
\caption{ Comparison of dynamically predicted polarized gluon
distribution with the NNPDF bounds [1]. } \label{fig10}
\end{figure}

\newpage
\begin{figure}[htp]
\centering

\vskip -3cm
\includegraphics[width=0.6\textwidth]{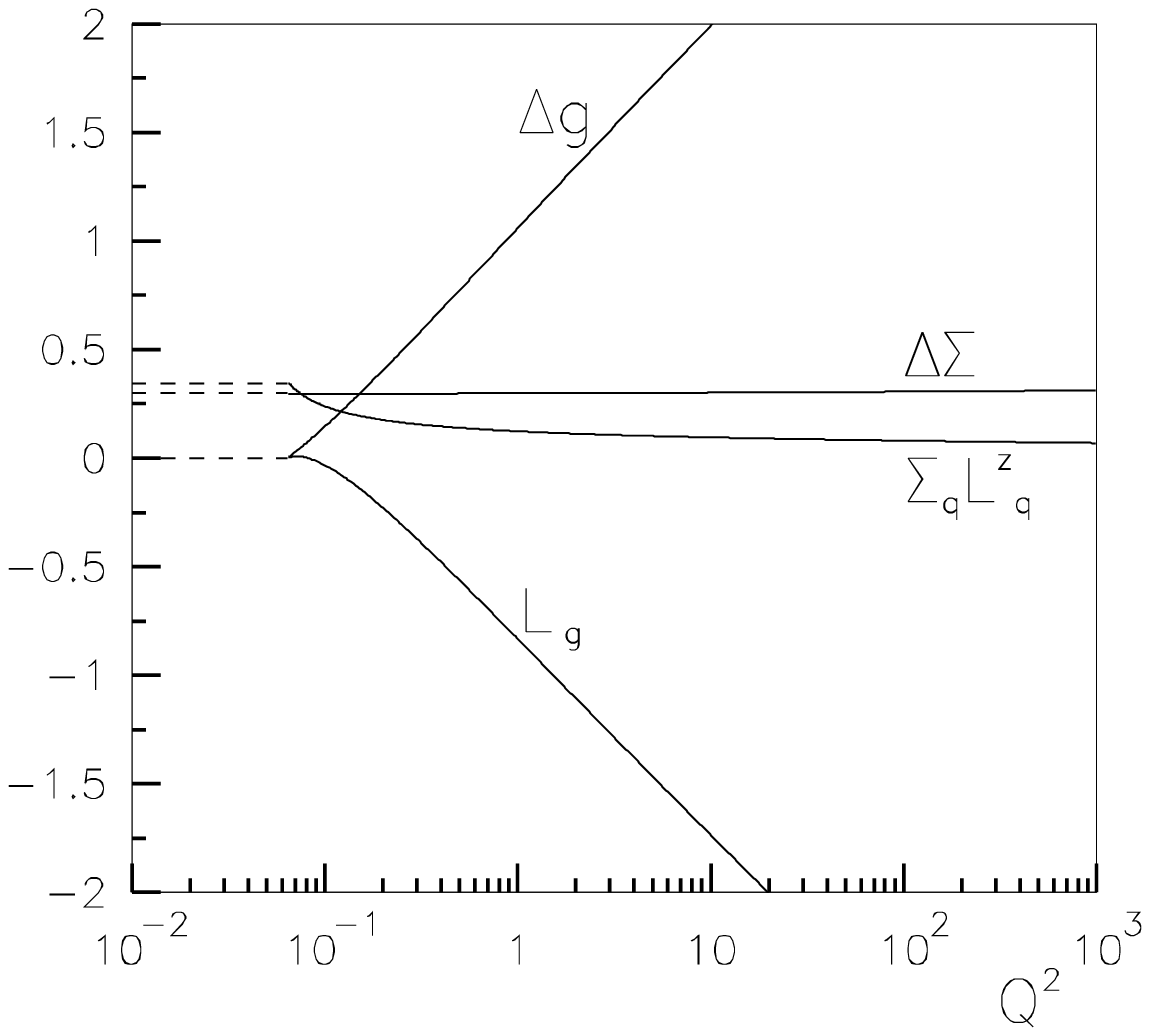}

\vskip -3cm
\caption{Contributions of spin and orbital motion of the partons to
the proton spin and their evolutions with $Q^2$. } \label{fig11}

\centering

\includegraphics[width=1.2\textwidth]{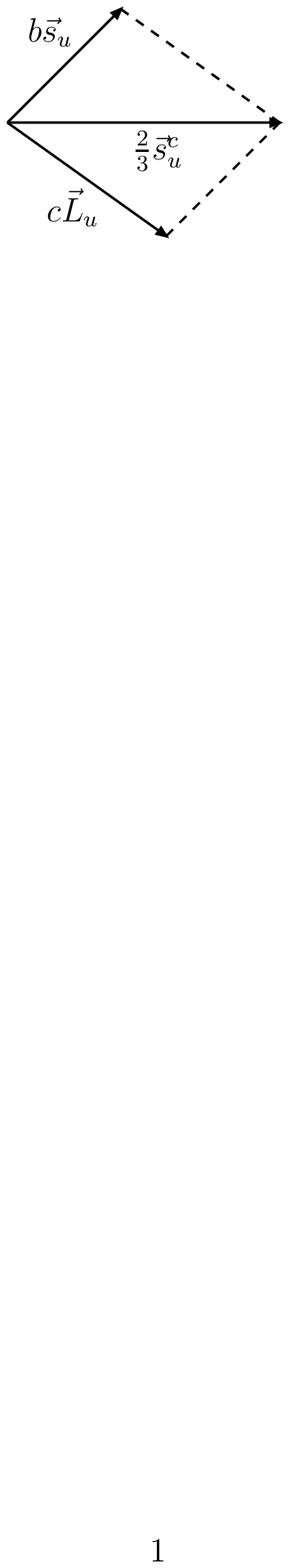}

\vskip -10cm

\vskip -6.0cm \caption{ A schematic diagram of the proton spin
crisis: Orbital angular momentum $\vec{L}_u$ of a valence u-quark at
a bound state scale $\mu^2$ impels the direction of the u-quark spin
($\vec{s}_u$) to deviate the polarized direction of the proton
($\vec{s}^c_u$) and gives $\Delta\Sigma<1$.
 } \label{fig12}
\end{figure}

\newpage
\begin{figure}[htp]

\vskip 3cm \centering

\vskip -3cm
\includegraphics[width=0.8\textwidth]{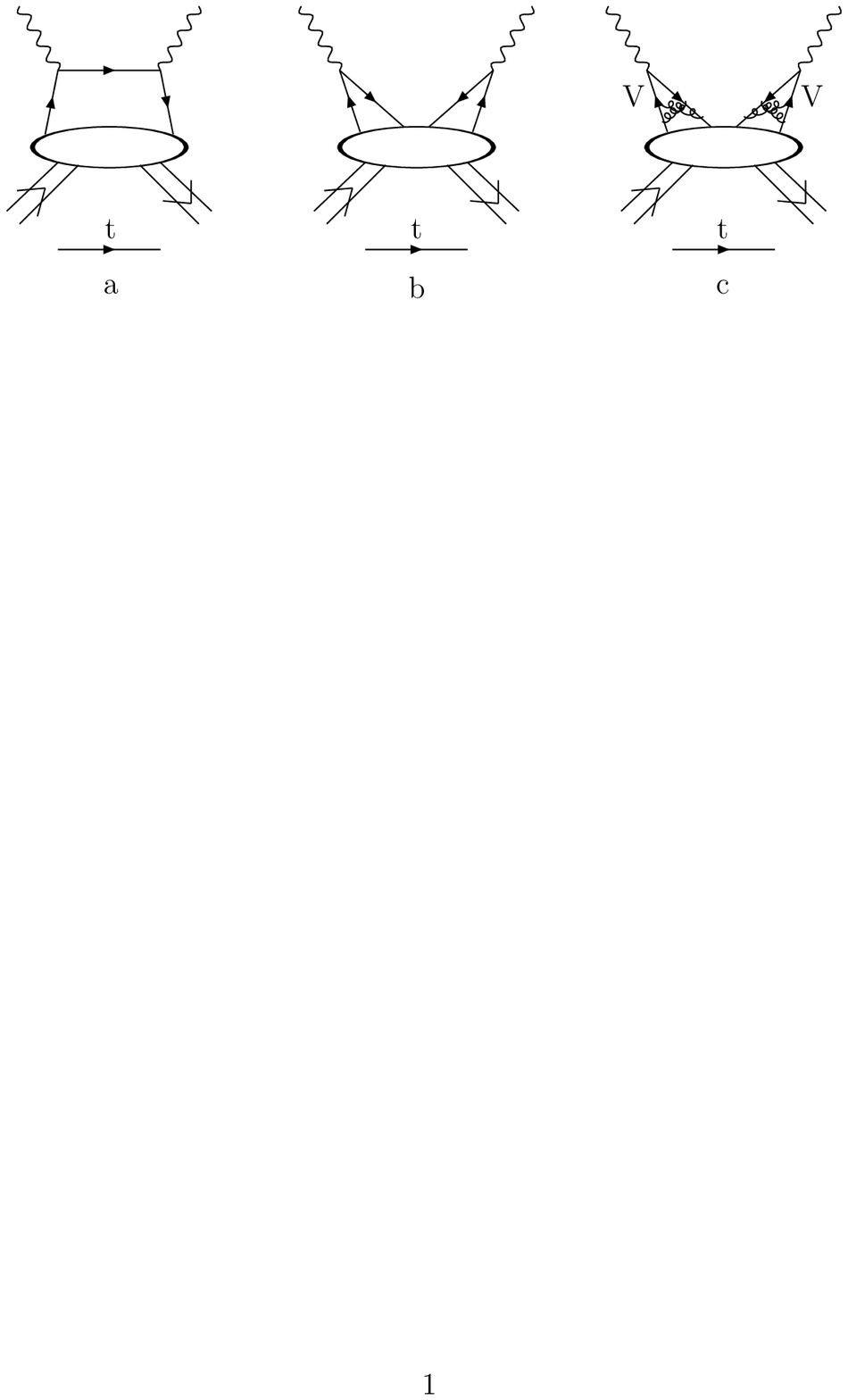}

\vskip -8cm
\caption{ The time ordered decomposing of DIS diagrams.
(a) The struck quarks are on-mass-shell since they have only forward
component. (b) A "cat ear" diagram, which vanishes in the collinear
factorization schema. (c) The "cat ear" diagram with higher order
QCD corrections, which are non-vanished at low $Q^2$, but can be
isolated using a naive VMD model.} \label{fig13}
\centering

\vskip -1cm
\includegraphics[width=0.6\textwidth]{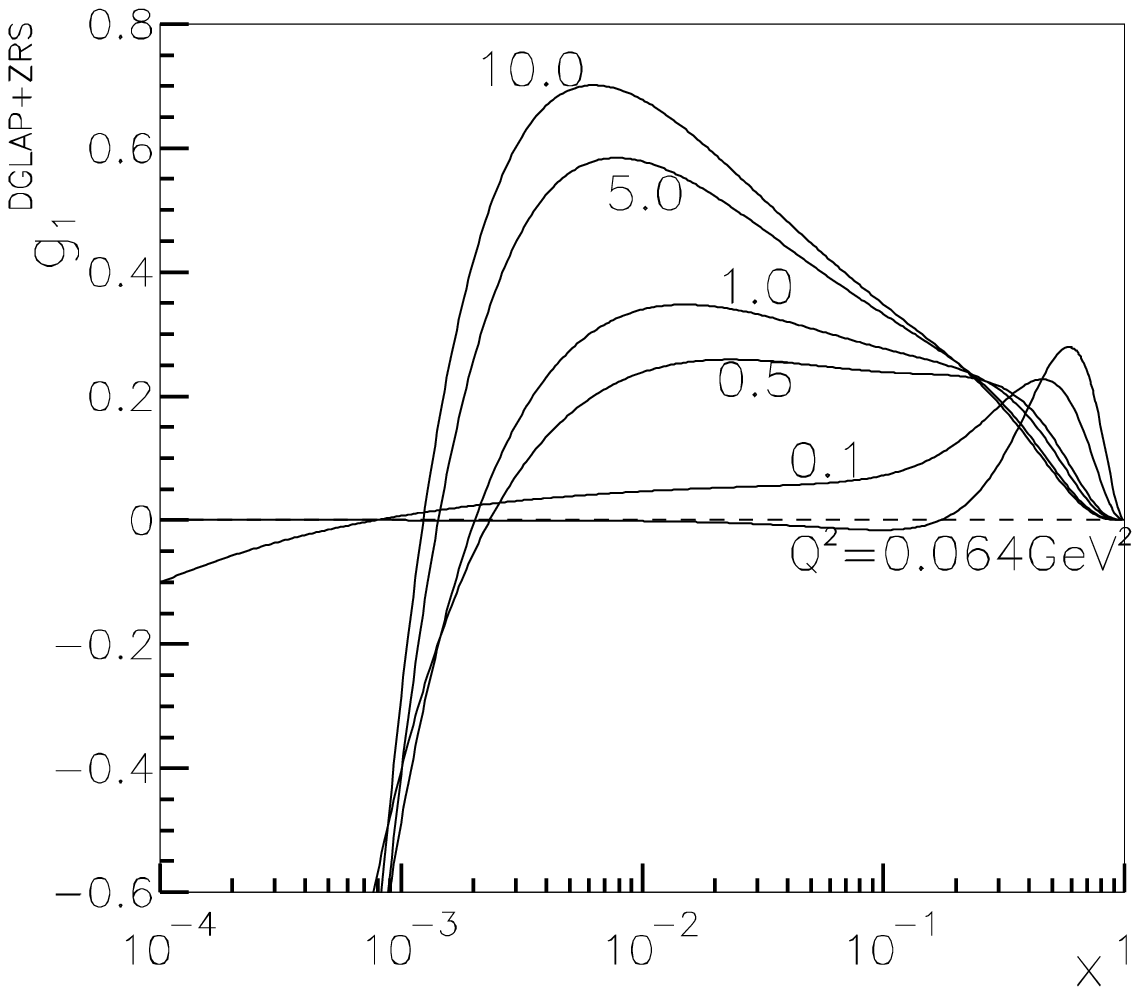}

\vskip -3cm
\caption{Perturbative $g^{DGLAP+ZRS}_1$. All partons are evolved
from three valence quarks at $\mu^2=0.064GeV^2$.} \label{fig14}
\end{figure}

\newpage
\begin{figure}[htp]
\centering

\vskip -3cm
\includegraphics[width=0.6\textwidth]{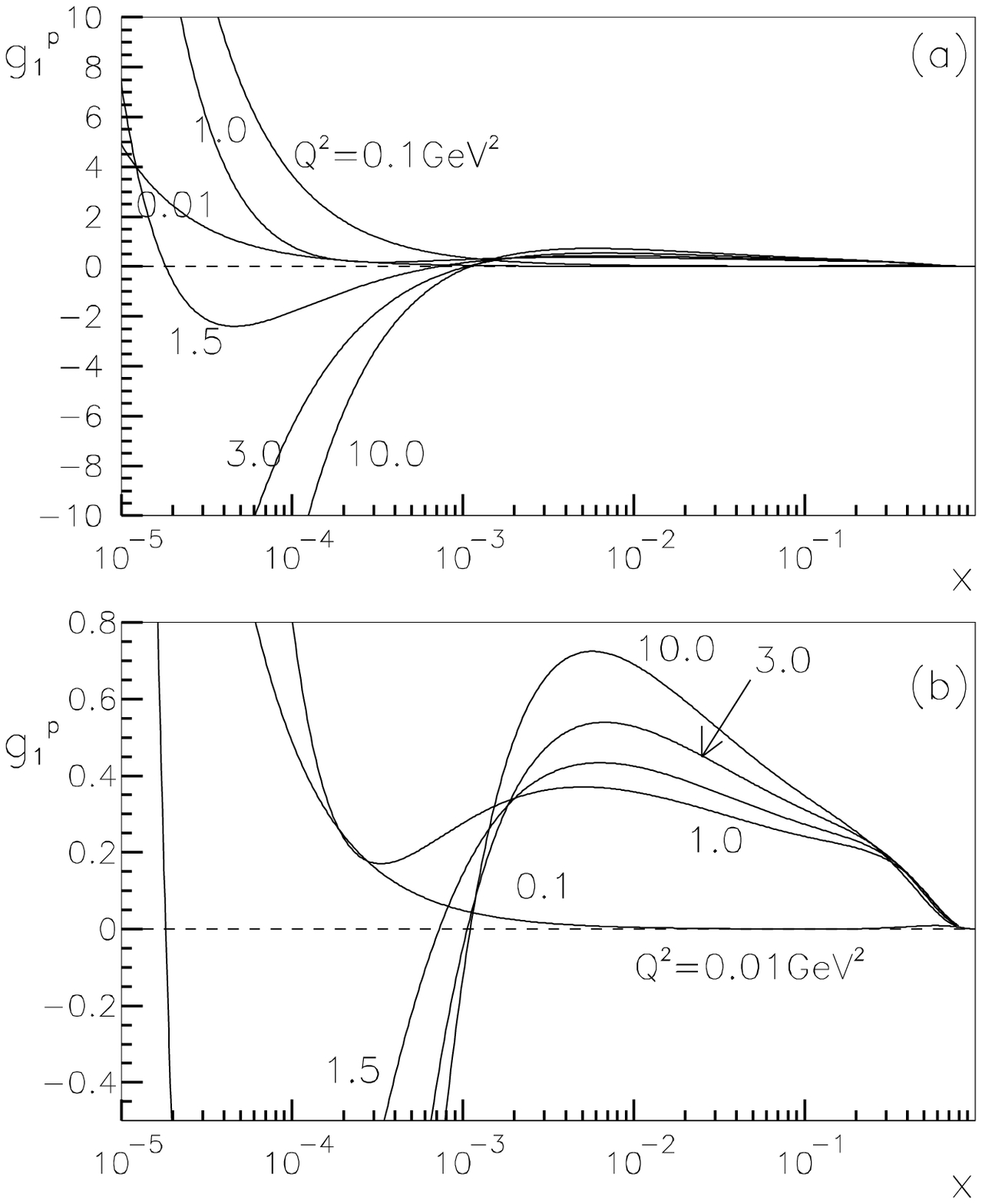}

\vskip -3cm
\caption{$g_1^p$ evolutions at different values of $Q^2$ in (a)
large and (b) small scales.} \label{fig15}

\centering

\includegraphics[width=0.6\textwidth]{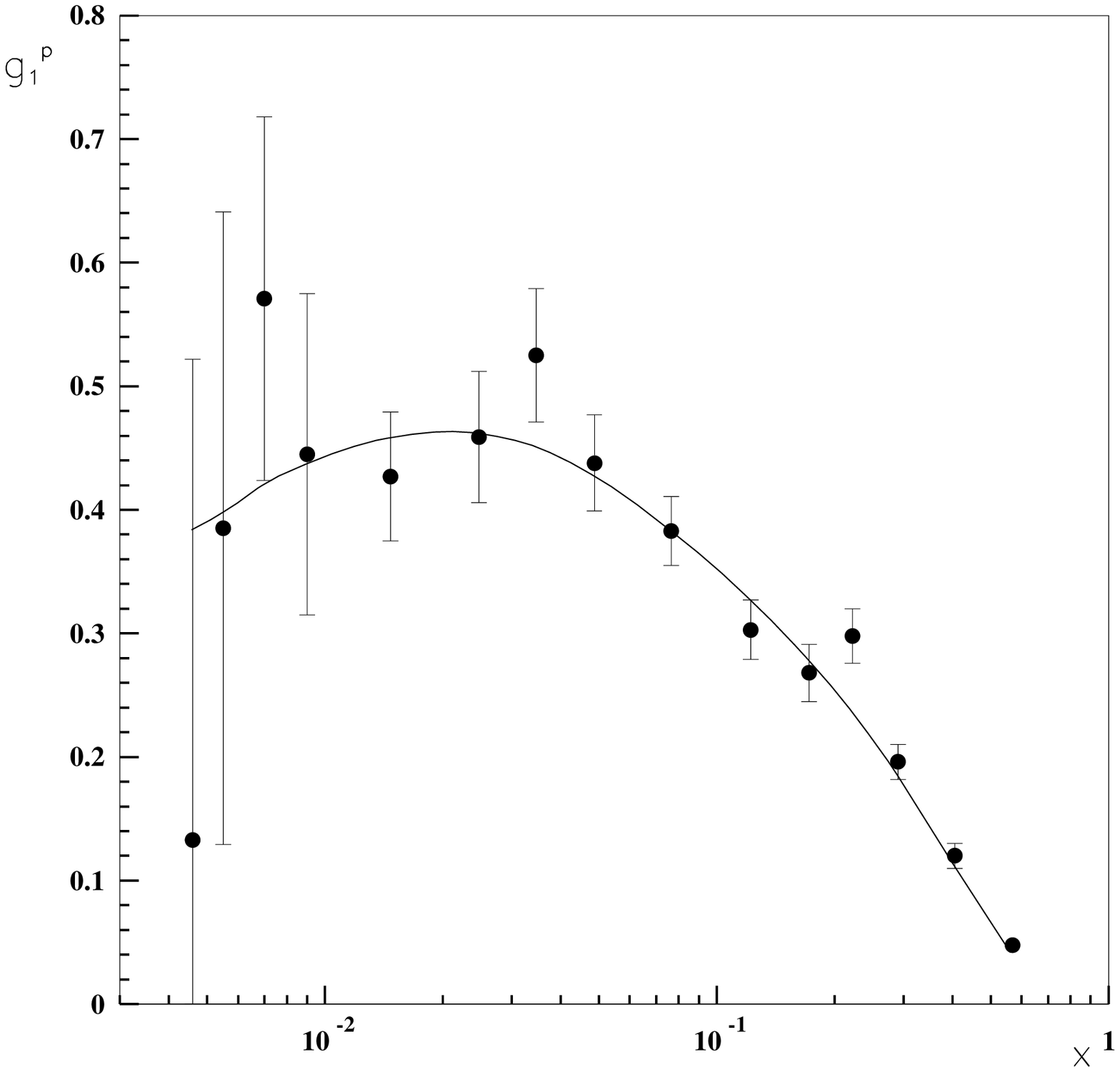}
\vskip -3cm
\caption{Predicted $g^p_1$ at $x>10^{-3}$ and comparisons with the
COMPASS data [47]. Note that the values of $Q^2(x)$ of each measured
point are different (see Table I of Ref.[29]).} \label{fig16}
\end{figure}

\newpage
\begin{figure}[htp]
\centering

\vskip -3cm
\includegraphics[width=0.6\textwidth]{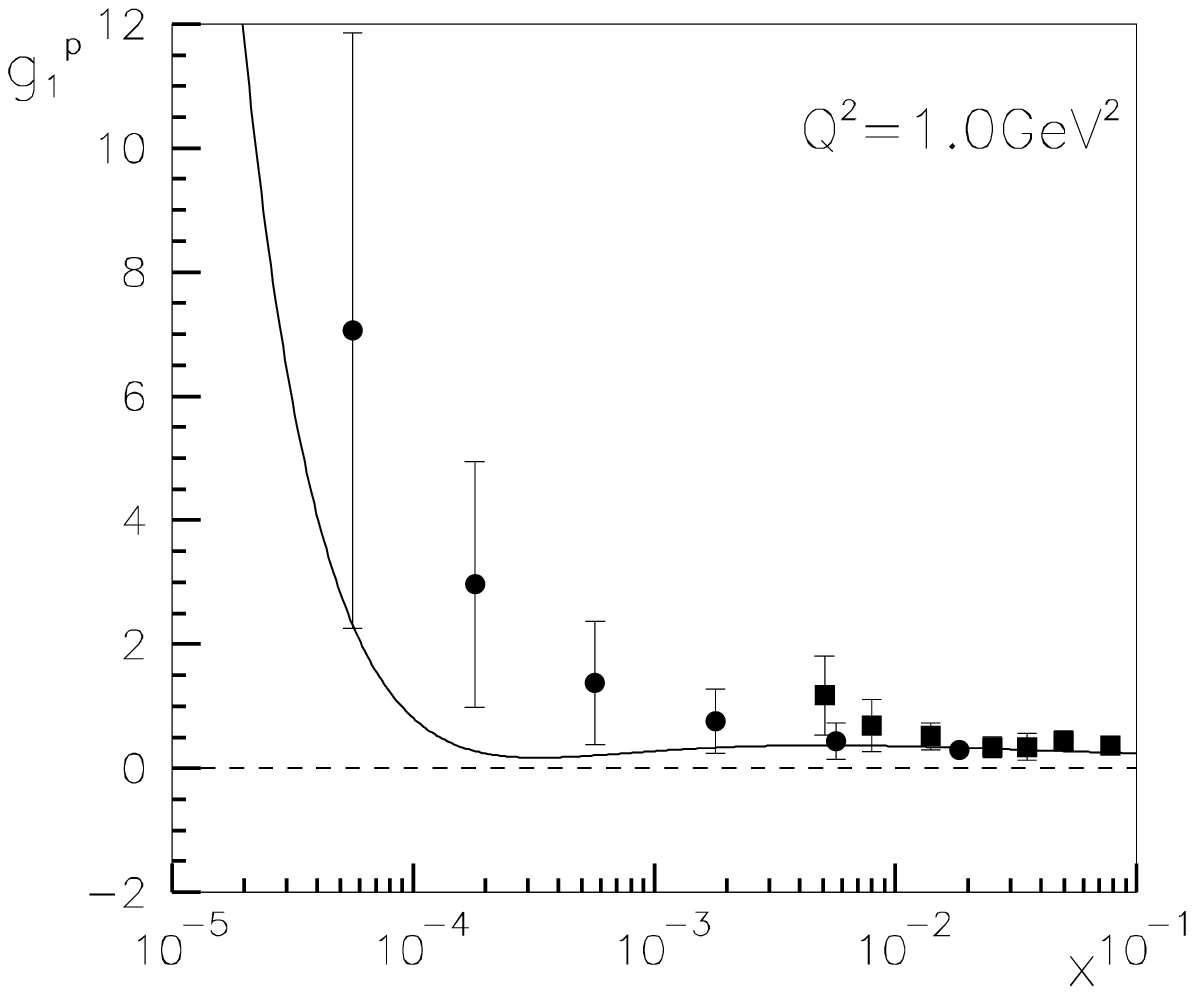}

\vskip -3cm
\caption{Predicted $g_1^p$ at $Q^2=1GeV^2$ and the comparison with
the HERA data [48].} \label{fig17}
\centering

\vskip -1cm
\includegraphics[width=0.6\textwidth]{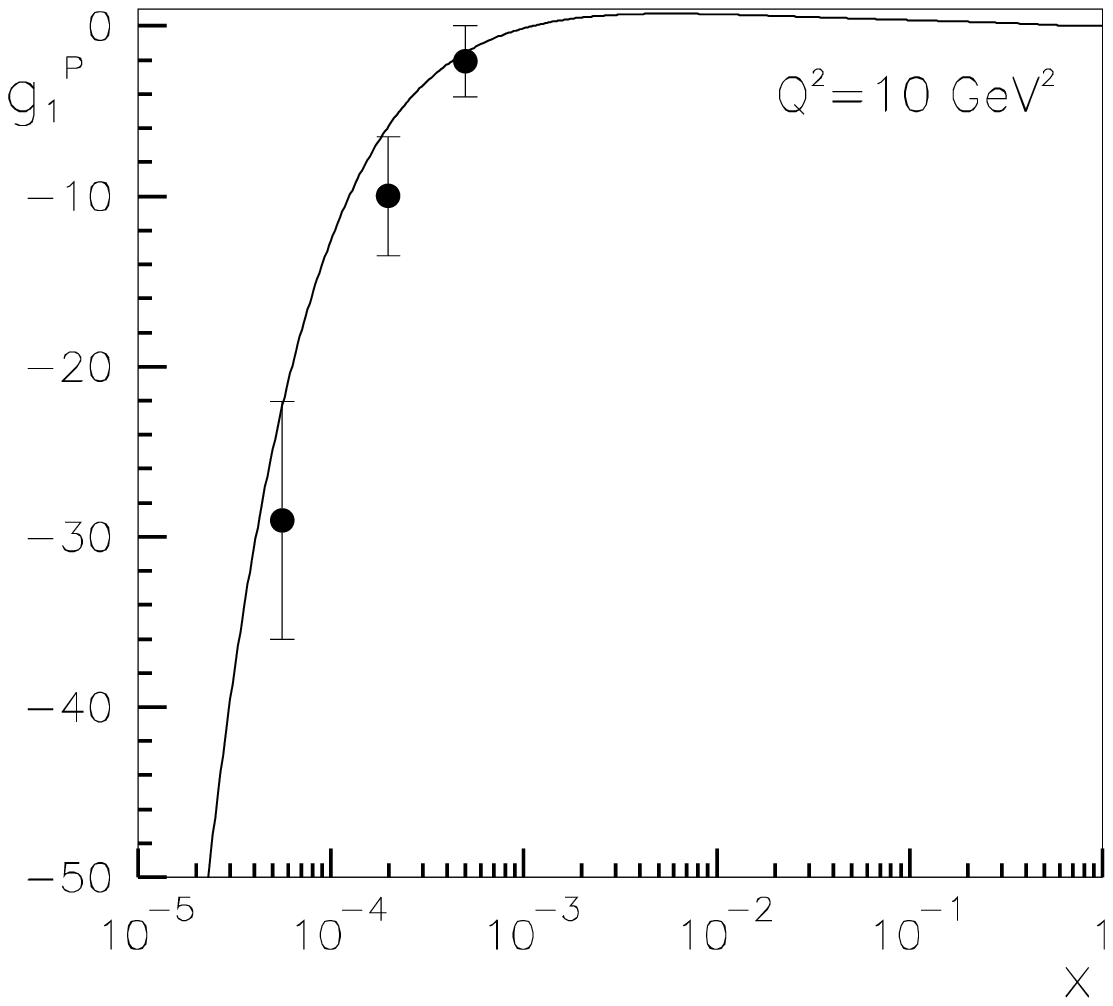}

\vskip -3cm
\caption{Predicted $g_1^p$ at $Q^2=10GeV^2$ and the comparison with
the HERA "data", which are based on the NLO QCD predictions with the
statistical errors expected at HERA [49].} \label{fig18}
\end{figure}

\newpage
\begin{figure}[htp]
\centering

\vskip -3cm
\includegraphics[width=0.6\textwidth]{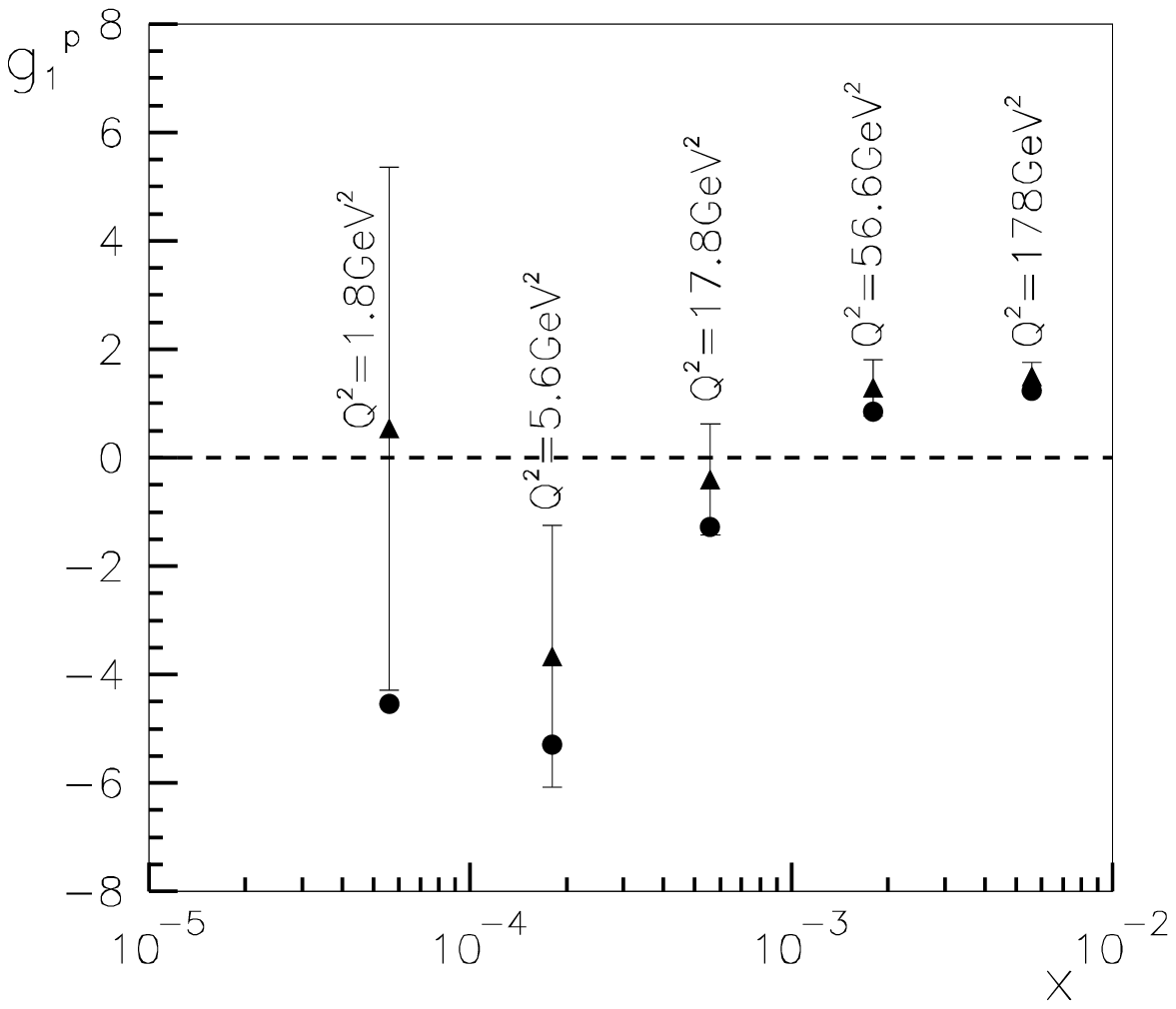}

\vskip -3cm
\caption{Predicted $g_1^p$ at $Q^2=1.8GeV^2$, $5.6GeV^2$ and
$16.5GeV^2$ at $x<10^{-3}$ (circles) and the comparison with the
HERA data (triangles).} \label{fig19}
%\end{figure}

%\newpage
%\begin{figure}[htp]
\centering

\includegraphics[width=0.6\textwidth]{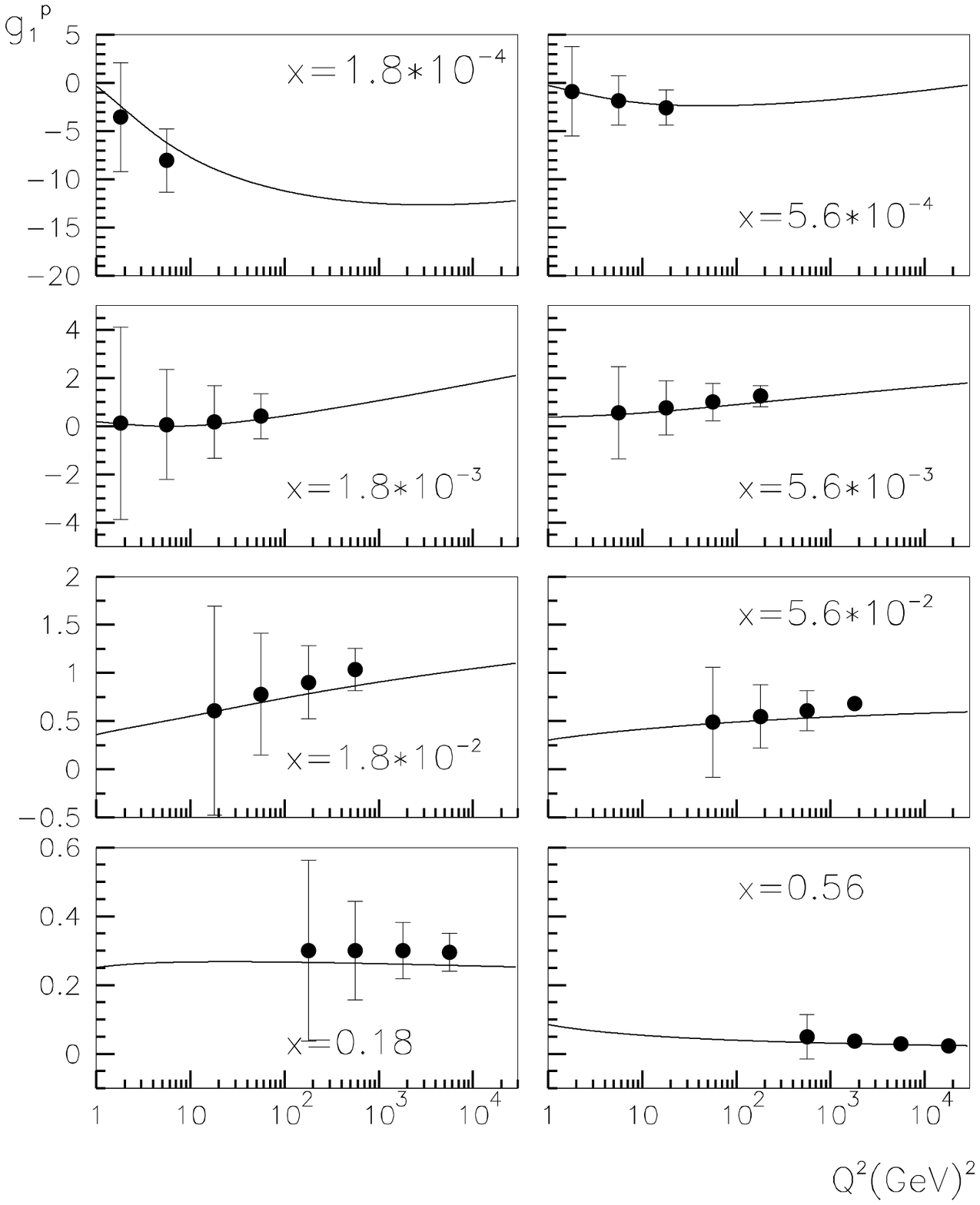}

\vskip -3cm
\caption{Predicted $Q^2$-dependence of $g_1^p$ with fixed values of
$x$. The data are taken from [50]. } \label{fig20}
\end{figure}

\newpage
\begin{figure}[htp]
\centering

\vskip -3cm
\includegraphics[width=0.6\textwidth]{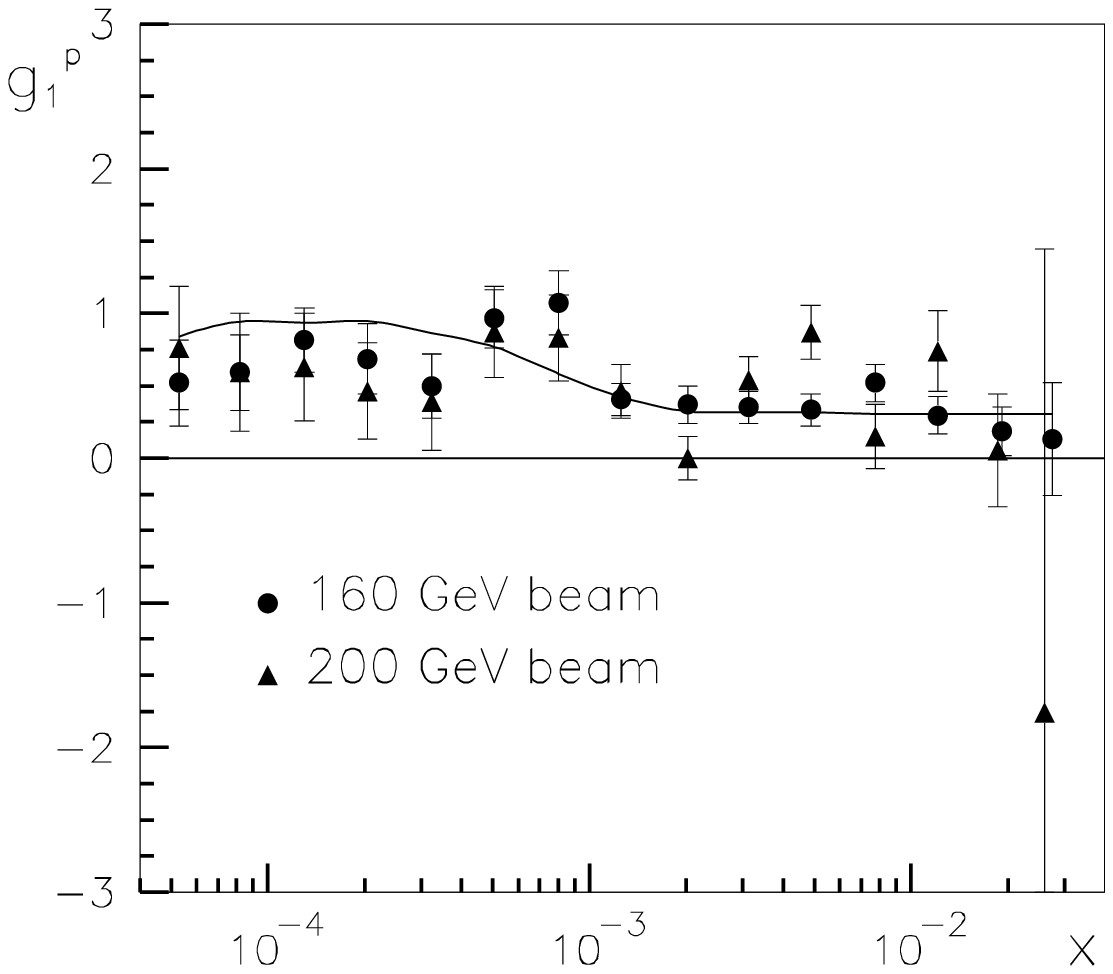}

\vskip -3cm
\caption{Predicted $g^p_1$ as a function of $x$ with different
measured $Q^2(x)$ (solid curve). Note that the low values of $x$
connect with the low values of $Q^2(x)$. The data are taken from
COMPASS primary results with two different beam energies [14,15,16,17,18]. }
\label{fig21}

\centering

\vskip -1cm
\includegraphics[width=0.6\textwidth]{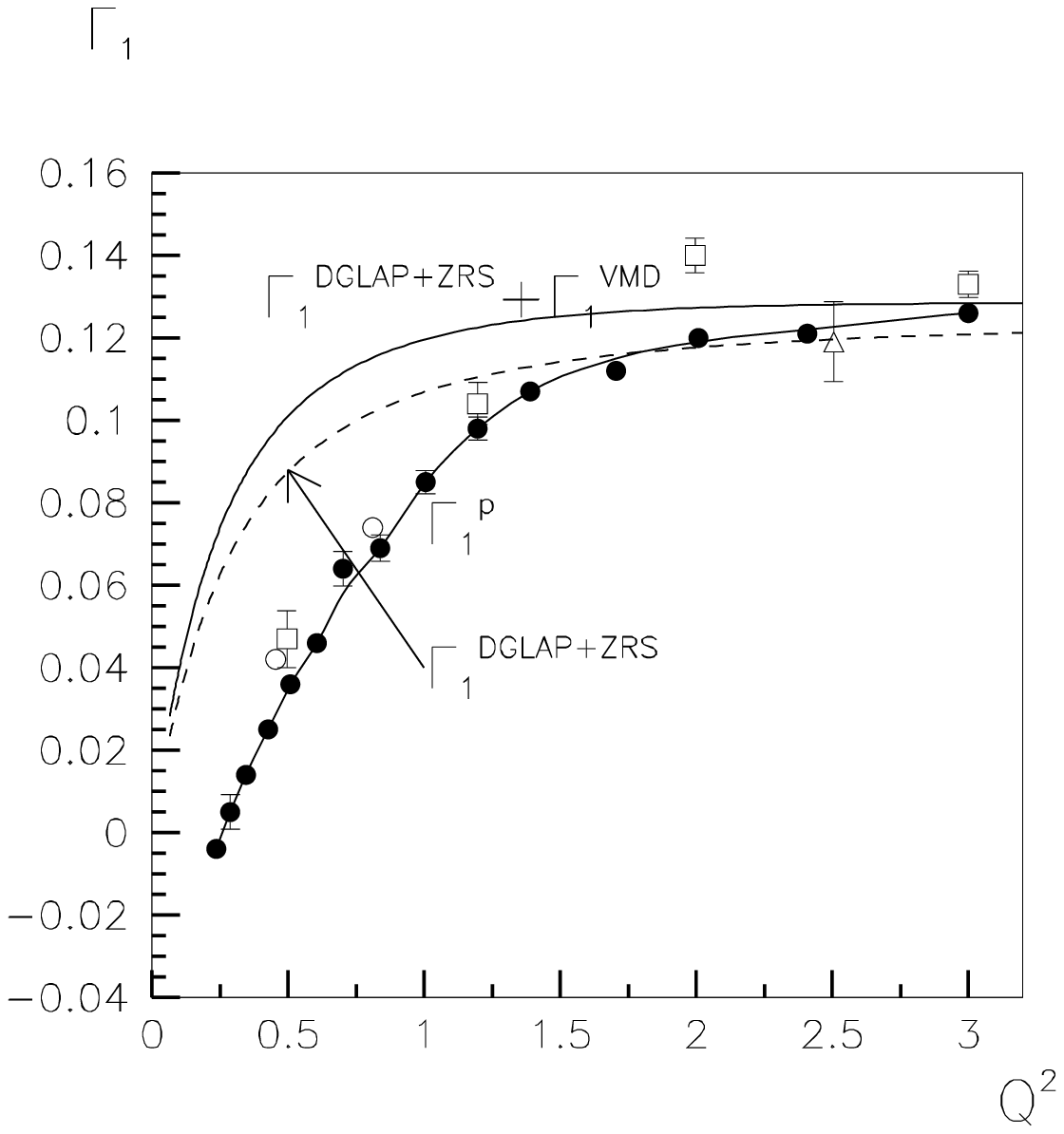}

\vskip -3cm
\caption{ Contribution of quark helicity
$\Gamma_1^{DGLAP+ZRS}(Q^2)$ (dashed curve) and combining VMD
contribution $\Gamma_1^{DGLAP+ZRS}(Q^2)+\Gamma_1^{VMD}(Q^2)$ (solid
curve). The data are taken from Hermes experiment at DESY [92,93], the
E143 experiment at SLAC [94] and the EG1a experiment using the CLAS
detector at JLab [95,96,97,98].} \label{fig22}
\end{figure}

\newpage
\begin{figure}[htp]
\centering

\vskip -3cm
\includegraphics[width=0.6\textwidth]{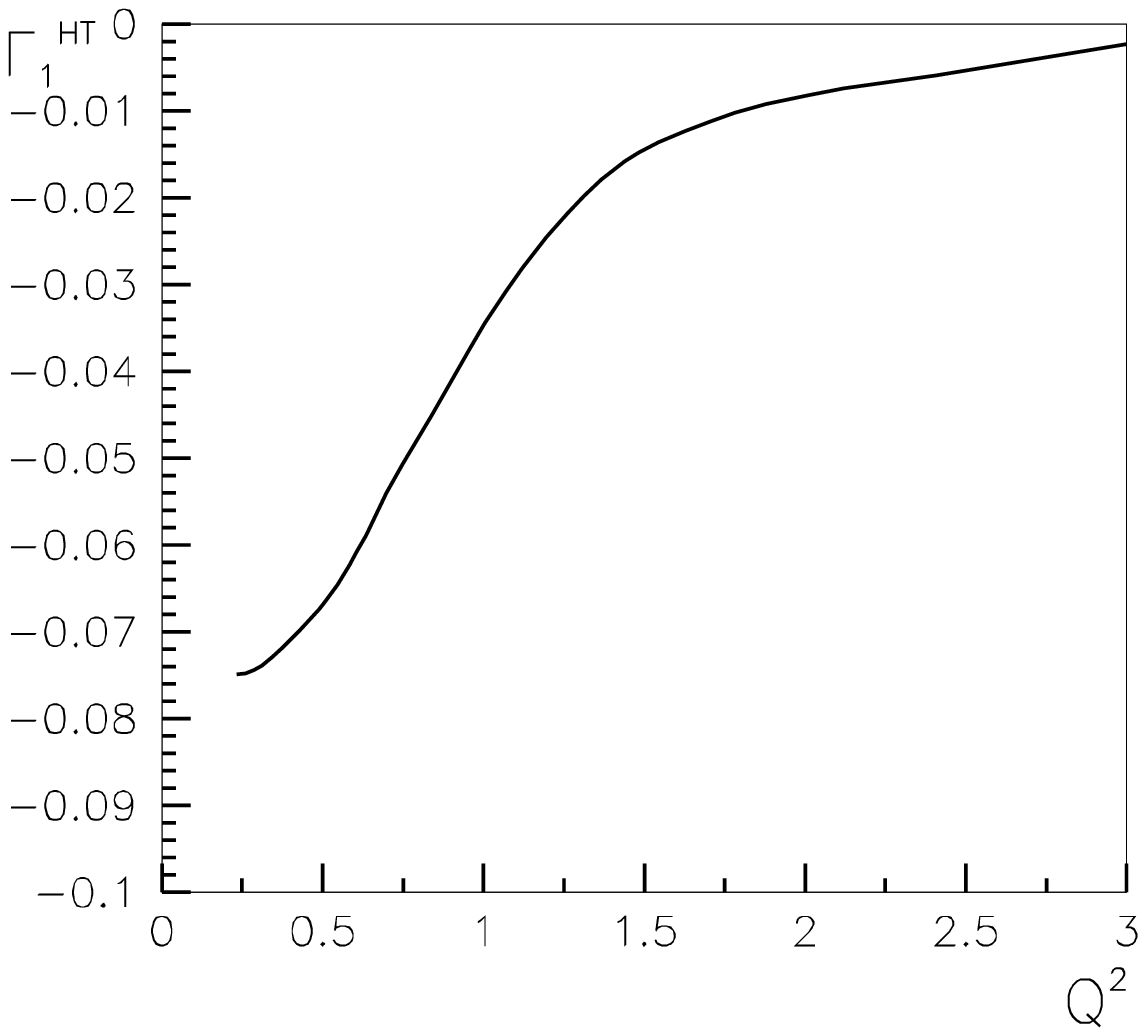}

\vskip -3cm
\caption{ Contribution of higher twist
$\Gamma_1^{HT}(Q^2)$ (smoothed curve) is taken from
data-[$\Gamma_1^{DGLAP+ZRS}(Q^2)+\Gamma_1^{VMD}(Q^2)]$.}

\centering

\vskip -1cm
\includegraphics[width=0.6\textwidth]{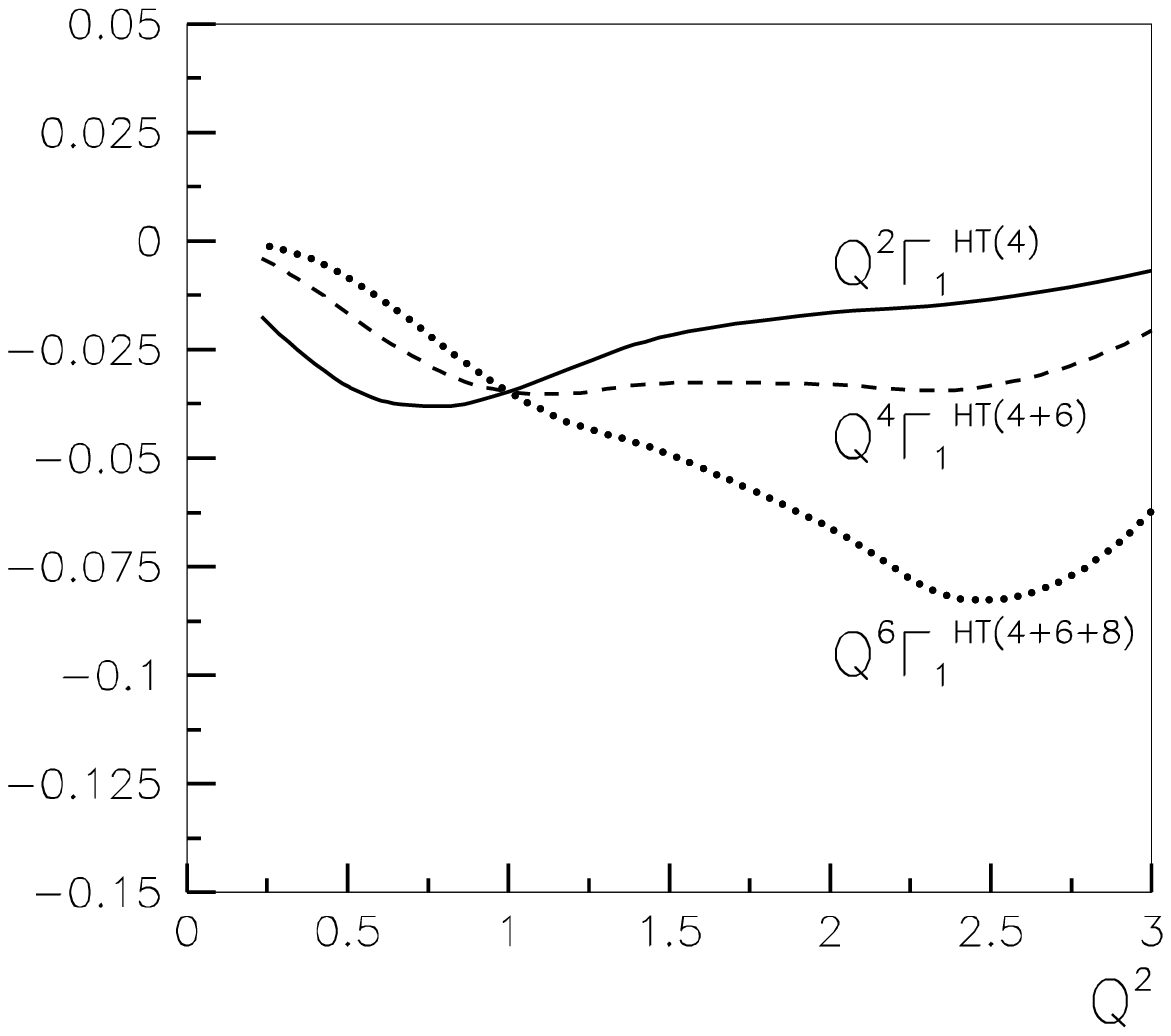}

\vskip -3cm
\caption{ Three different analysis of the higher twist
contributions.} \label{fig24}
\end{figure}

\newpage
\begin{figure}[htp]
\centering

\vskip -3cm
\includegraphics[width=0.7\textwidth]{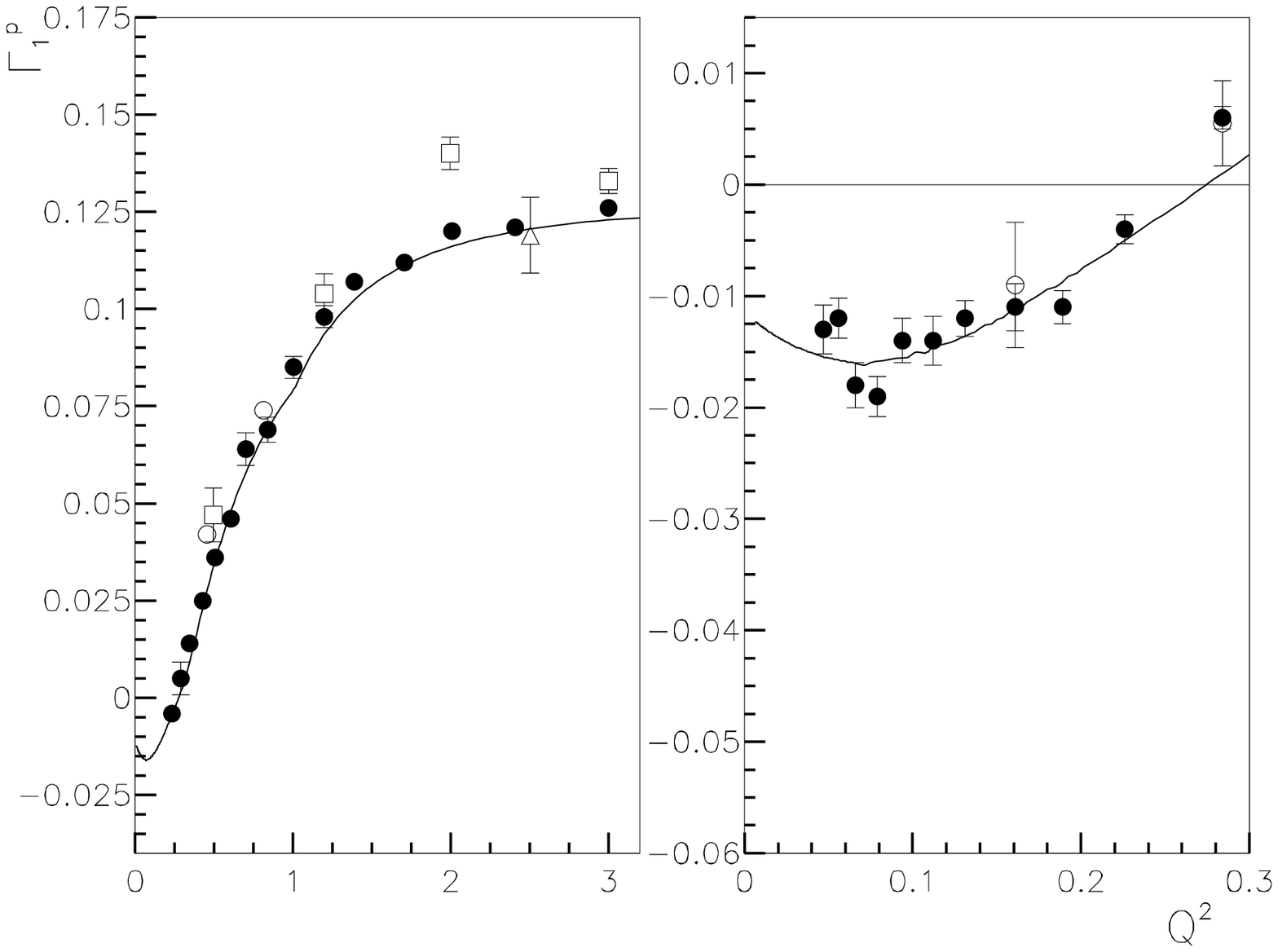}

\vskip -3cm
\caption{The $Q^2$ dependence of $\Gamma_1^p(Q^2)$
calculated by Eqs. (4.2.6) and (4.2.7). The data are taken from
[92,93,94,95,96,97,98].} \label{fig25}

\centering

\vskip -1cm
\includegraphics[width=0.7\textwidth]{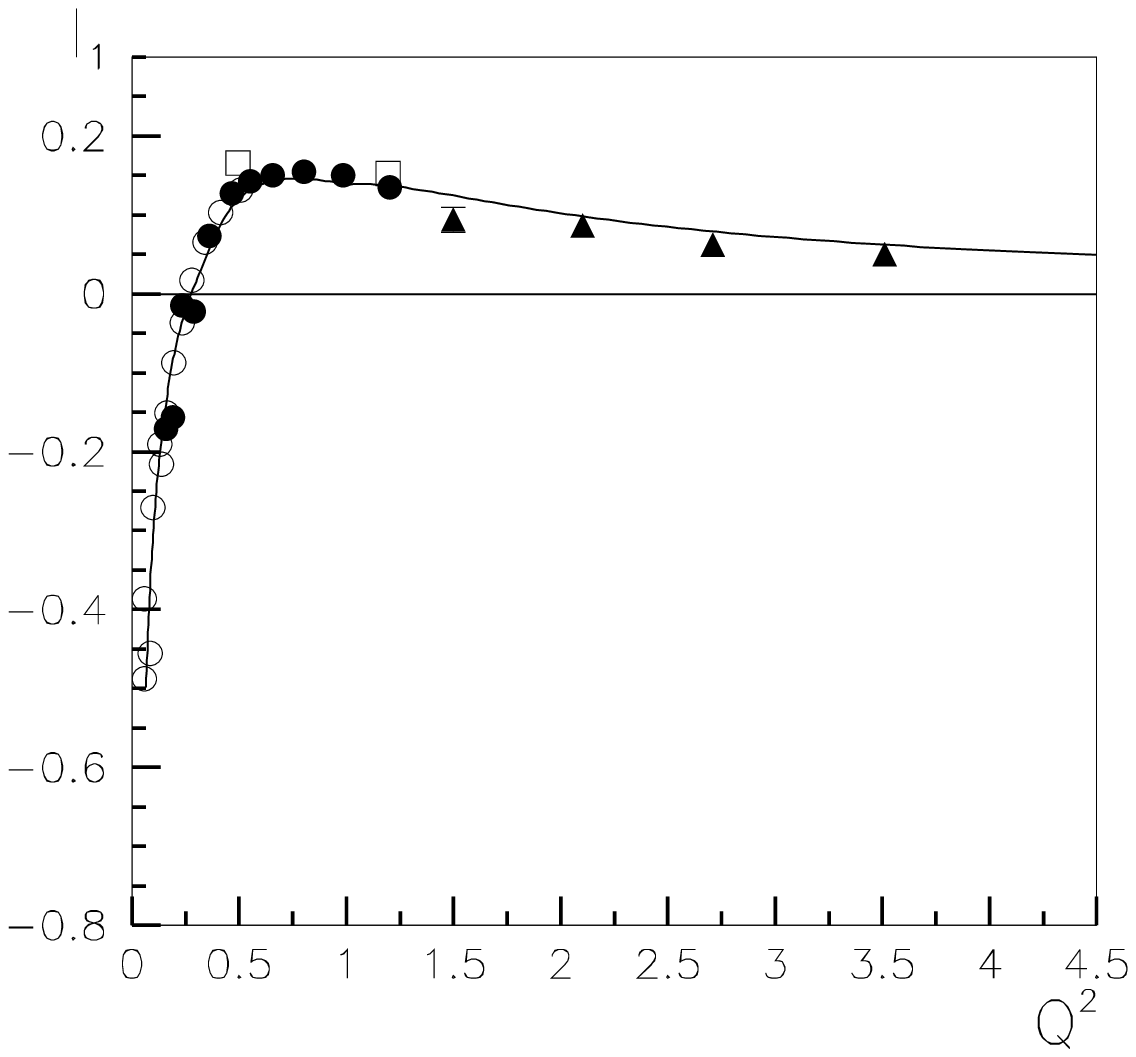}

\vskip -3cm
\caption{ The $Q^2$ dependence of $I_1^p(Q^2)$ calculated
by Eq. (4.1.3). The data are taken from [92,93,94,95,96,97,98].} \label{fig26}
\end{figure}


\newpage
\begin{thebibliography}{99}

\bibitem{1} NNPDF Collaboration, E. R. Nocera, R. D. Ball, S. Forte,
G. Ridolfi and J. Rojo, to be published in Nucl.Phys. B, hep-ph/
1406.5539.

\bibitem{2} G. Parisi and R. Petronzio, Phys. Lett. $\bf{B62}$, 331 (1976).

\bibitem{3}V.A. Novikov, M.A. Shifman, A.I. Vainshtein, V.I. Zakharov, JETP
Lett., $\bf{24}$, 341 (1976).

\bibitem{4} M. Gl$\ddot{u}$ck, E. Reya, Nucl.Phys. $\bf{B 130}$, 76 (1977).

\bibitem{5} X.R. Chen, J.H. Ruan, R. Wang, P.M. Zhang and W. Zhu, Int.
J. Mod. Phys. $\bf{E23}$, 14500057 (2014), hep-ph/1306.1872.

\bibitem{6} X.R. Chen, J.H. Ruan, R. Wang, P.M. Zhang and W. Zhu,
$\bf{E23}$, 1450058 (2014), hep-ph/1306.1874.

\bibitem{7} G. Altarelli, G. Parisi, Nucl. Phys. $\bf{B126}$, 298 (1977).

\bibitem{8} V.N. Gribov, L.N. Lipatov, Sov. J. Nucl. Phys. $\bf{15}$, 438
(1972).

\bibitem{9} Yu.L. Dokshitzer, Sov. Phys. JETP $\bf{46}$, 641 (1977).

\bibitem{10} W. Zhu, Nucl. Phys. $\bf{B551}$, 245 (1999), hep-ph/9809391;

\bibitem{11} W. Zhu, J.H. Ruan, Nucl. Phys. $\bf{B559}$, 378 (1999),
hep-ph/9907330v2.

\bibitem{12} W. Zhu and Z.Q. Shen, HEP $\&$ NP, $\bf{29}$, 109
(2005), hep-ph/0406213v3.

\bibitem{13} W. Zhu, Z.Q. Shen and J.H. Ruan, Nucl.Phys. $\bf{B692}$, 417 (2004), hep-ph/0406212v2[hep-ph].


\bibitem{14} A.S. Nunes (on behalf of the COMPASS Collab.,)
Longitudinal double spin asymmetry $A^p_1$ and spin-dependent
structure function $g^p_1$ of the proton at low $x$ and low $Q^2$
from COMPASS,  Proceedings of the XV workshop on hihg energy spin
physics, Dubna, Russia, (2013), hep-ex/1405.5811.

\bibitem{15} COMPASS Collaboration, P. Abbon et al., Nucl. Instr. and Meth.
$\bf{A 577}$, 455 (2007).

\bibitem{16} E.S. Ageev et al., Phys. Lett.
$\bf{B612}$, 154 (2005).

\bibitem{17} V.Yu. Alexakhin et al., Phys. Lett.
$\bf{B647}$, 8 (2007).

\bibitem{18} M.G. Alekseev et al., Phys. Lett.
$\bf{B690}$, 466 (2010).

\bibitem{19} S.B. Gerasimov, Sov. J. Nucl. Phys. $\bf{2}$, 430 (1966).

\bibitem{20} S.D. Drell and A.C. Hearn, Phys. Rev. Lett. $\bf{16}$, 908
(1966).

\bibitem{21} NM Collaboration, J. Ashman et al., Nucl. Phys. $\bf{B328}$, 1 (1989) and references
therein.

\bibitem{22} M. Gl$\ddot{u}$ck, E. Reya, and A. Vogt, Eur. Phys.
J. $\bf{C5}$, 461 (1998)..

\bibitem{23} COMPASS Collaboration, C. Adolph, et. al.,
 hep-ex/1202.4064.

\bibitem{24} D. de Florian, R. Sassot, M. Stratmann and W. Vogelsang,
Phys. Rev. Lett. $\bf{113}$, 012001 (2014), hep-ph/1404.4293.

\bibitem{25} P. Djawotho for the STAR Collaboration, nucl-ex/1303.0543.

\bibitem{26} PHENIX Collaboration, A. Adare et al., Phys. Rev. $\bf{D90}$, 012007 (2014), hep-ex/1402.6296.

\bibitem{27} J. Ellis and M. Karliner, Phys. Lett., $\bf{B341}$, 397 (1995).

\bibitem{28} A. Airapetian, et.al., Phys, Pev. $\bf{D75}$, 012007 (2007).

\bibitem{29} B.Q. Ma, J. Phys, $\bf{G17}$, L53 (1991).

\bibitem{30} B.Q. Ma and S.J. Brodsky, The spin and flavor content of intrinisic
sea quarks, hep-ph/9707408.

\bibitem{31} B.Q. Ma, I. Schmidt and J. Soffer, Phys.
Lett. $\bf{B441}$, 461 (1998).

\bibitem{32} B.Q. Ma and I. Schmit, Phys. Rev.
$\bf{D58}$ 096008 (1998).

\bibitem{33} H.J. Melosh, Phys. Rev. $\bf{D9}$, 1095 (1974); E.
Wigner, Ann. Math. $\bf{40}$, 149 (1939).

\bibitem{34} X.D. Ji, J. Tang and P. Hoodbhoy, Phys. Rev.
Lett.$\bf{76}$, 740 (1996).

\bibitem{35} T.C. Meng, J.C. Pan, Q.B. Xie and W. Zhu, Phys. Rev.
$\bf{D40}$, 769 (1989).

\bibitem{36}  T.C. Meng, Invited talk given at the
Workshop on the Prospects of Spin Physics at HERA, DESY Zeuthen,
August 28-31, 1995 hep-ph/9510336.

\bibitem{37} J.C. Collins, D.E. Soper, G. Sterman, in: A.H. Mueller (Ed.),
Perturbative Quantum Chromodynamics, World Scientific, Singapore,
1989, p. 1.

\bibitem{38} W. Zhu, H.W. Xiong, J.H. Ruan, Phys. Rev. $\bf{D60}$, 094006
(1999).

\bibitem{39} W. Zhu, Nucl. Phys. $\bf{A753}$, 206 (2005).

\bibitem{40} J.J. Sakurai, currents and mesons, university of Chigag, Chigago (1969).

\bibitem{41} T. H. Bauer et al.,Rev. Mod. Phys. $\bf{50}$, 261 (1978).

\bibitem{42} G. Grammer Jr and J. D.Sullivan, in Electromagnetic Interactions of Hadrons, edited by A.
Donnachie and G. Shaw, Plenum, New York, 1978, Vol.2.

\bibitem{43} B. Badelek, J, Kwieci¨½ski, B. Ziaja, Eur. Phys. J.
$\bf{C26}$, 45 (2002).

\bibitem{44} B. Badelek, J, Kwieci¨½ski, B. Ziaja,Acta Phys. Polon. $\bf{B33}$, 3701 (2002).

\bibitem{45} P.D.B. Collins, An Introduction to Regge Theory and High Energy Physics, Cam-
bridge University Press, Cambridge, 1977.

\bibitem{46} S.J. Brodsky and G. Farrar, Phys. Rev. Lett. $\bf{31}$, 1153 (1973).

\bibitem{47} COMPASS Collaboration, M.G. Alekseev, et al
Phys. Lett. $\bf{B690}$, 466 (2010).

\bibitem{48} R. D. Ball, A. Deshpande, S. Forte, V. W. Hughes, J. Lichtenstadt,, G.
Ridolf, Measurement of the polarized sreucture function
$g_1^p(x,Q^2)$ at HERA,  hep-ph/9609515.

\bibitem{49} J. Kwiecinski and B. Ziaja, hep-ph/9802386.

\bibitem{50} A. De Roeck, A. Deshpande, V.W. Hughes,
J. Lichtenstadt, G. Radel Eur. Phys. J. $\bf{C6}$, 121 (1999).

\bibitem{51} E.C. Aschenauer, at. al., eRHIC Design Study: An Electron-Ion Collider at BNL, hep-ph/1409.1633.

\bibitem{52} X.R. Chen, An Electro Ion Collider Plan in China, Invited talk at the
21st International Symposium on Spin Physics, Beijing, China, Oct.
20-24 (2014).


\bibitem{53} L.N. Lipatov, Sov. J. Nucl. Phys. $\bf{23}$, 338 (1976).


\bibitem{54} V.S. Fadin, E.A. Kuraev, L.N. Lipatov, Phys. Lett.$\bf{B 60}$, 50
(1975).

\bibitem{55}  E.A. Kuraev, L.N. Lipatov, V.S. Fadin, Sov. Phys.
JETP$\bf{44}$, 443 (1976).

\bibitem{56} E.A. Kuraev, L.N. Lipatov, V.S. Fadin,
Sov. Phys. JETP $\bf{45}$, 199 (1977).

\bibitem{57} I.I. Balitsky, L.N. Lipatov,
Sov. J. Nucl. Phys. $\bf{28}$, 822 (1978).

\bibitem{58} I.I. Balitsky, L.N.
Lipatov, JETP Lett. $\bf{30}$, 355 (1979) .

\bibitem{59}
I. Balitsky, Nucl. Phys. $\bf{B463}$ (1996) 99.

\bibitem{60} Yu. Kovchegov, Phys.
Rev. $\bf{D60}$ (1999) 034008.

\bibitem{61} Yu. Kovchegov, Phys. Rev. $\bf{D61}$
(2000) 074018.

\bibitem{62}
J. Jalilian-Marian, A. Kovner, L. McLerran, and H. Weigert, Phys.
Rev. $\bf{D55}$, 5414 (1997).

\bibitem{63} J. Jalilian-Marian, A. Kovner, A.
Leonidov, and H. Weigert, Nucl. Phys. $\bf{B504}$, 415 (1997).

\bibitem{64} J. Jalilian-Marian, A. Kovner, A.
Leonidov, and H. Weigert,Phys.Rev. $\bf{D59}$, 014014 (1998).

\bibitem{65} H.Weigert, Nucl. Phys. $\bf{A703}$,823 (2002).

\bibitem{66} E. Iancu, A. Leonidiv, and L.McLerran, ibid.
$\bf{A692}$, 583 (2001).

\bibitem{67} E. Iancu, A. Leonidiv, and L.McLerran,Phys. Lett. $\bf{B510}$, 133 (2001).

\bibitem{68} M. Ciafaloni, Nucl. Phys. $\bf{B296}$, 49 (1988).

\bibitem{69} S. Catani, F.Fiorani, and G. Marchesini, Phys. Lett. $\bf{B234}$, 339 (1990).

\bibitem{70} S. Catani, F.Fiorani, and G. Marchesini,Nucl. Phys. $\bf{B336}$, 18 (1990).

\bibitem{71} G. Marchesini, Nucl. Phys.$\bf{B445}$, 49 (1995).

\bibitem{72} D. Kotlorz and A. Kotlorz,  Acta Phys. Polon. $\bf{B39}$, 1913 (2008).

\bibitem{73} B.I. Ermolaev, M. Greco, S.I. Troyan, Riv. Nuovo Cim. $\bf{33}$, 57
(2010).

\bibitem{74} L.N. Lipatov, Zh.Eksp.Teor.Fiz. $\bf{82}$, 991 (1982).

\bibitem{75} L.N. Lipatov,Phys. Lett.$\bf{B116}$, 411 (1982).

\bibitem{76} M. Gl$\ddot{u}$ck, E. Reya, and A. Vogt, Eur. Phys.
J. $\bf{C5}$, 461 (1998).

\bibitem{77} J.D. Bjorken, Phys. Rev. $\bf{148}$, 1467 (1966).

\bibitem{78} Particle Data Group, S. Eidelman et al., Phys. Lett. $\bf{B592}$, 1 (2004).

\bibitem{79} D. Drechsel, S.S. Kamalov and L. Tiator, Phys. Rev. $\bf{D63}$, 114010 (2001)
hep-ph/0008306.

\bibitem{80} M. Anselmino, B.L. Ioffe and E. Leader, 1989, Sov. J. Nucl.
Phys. $\bf{49}$, 136 (1989).

\bibitem{81} V.D. Burkert and B.L. Ioffe,  Phys.
Lett. $\bf{B296}$, 223 (1992).

\bibitem{82}  J. Soffer and O.V. Teryaev, Phys.
Rev. Lett. $\bf{70}$, 3373 (1993).

\bibitem{83} J. Soffer and O. Teryaev, Phys.
Rev. $\bf{D70}$, 116004 (2004).

\bibitem{84} D. Drechsel and L. Tiator Ann. Rev. Nucl. Part. Sci. $\bf{54}$, 69 (2004), nucl-th/0406059.

\bibitem{85} M. Gorchtein, D. Drechsel, M.M. Giannini, E. Santopinto and L.
Tiator, Phys. Rev. $\bf{C70}$, 055202 (2004) 055202, hep-ph/0404053.

\bibitem{86} B. Badelek, J. Kiryluk, and J. Kwiecinski Phys. Rev. $\bf{D61}$, 014009, hep-ph/9907569.

\bibitem{87} B. Badelek, J. Kwiecinski and B. Ziaja, Eur. Phys. J. $\bf{C26}$, 45
(2002), hep-ph/0206188.

\bibitem{88} D. Burkert and Z.J. Li, Phys. Rev. $\bf{D47}$, 46 (1993).

\bibitem{89} V. Bernard, Prog. Part. Nucl. Phys. $\bf{60}$, 82 (2006).

\bibitem{90} V. Bernard et al., Phys. Rev. $\bf{D67}$, 076008 (2003).

\bibitem{91} X. Ji et al., Phys. Lett. $\bf{B472}$, 1 (2000).

\bibitem{92} HERMES Collaboration, A. Airapetian et al., Eur. Phys. J. C26
(2003) 527, hep-ex/0210047.

\bibitem{93} HERMES Collaboration, A. Airapetian et al.,Phys. Rev., $\bf{D75}$, 012007 (2007).

\bibitem{94} E143 Collaboration, K. Abe et al., Phys. Rev. Lett. $\bf{78}$, 815 (1997),
hep-ex/9701004.

\bibitem{95} CLAS Collaboration, R. Fatemi et al., Phys. Rev. Lett. $\bf{91}$, 222002 (2003),
nucl-ex/0306019.

\bibitem{96}  K. V. Dharmawardane et al., Phys. Lett. $\bf{B
641}$, 11 (2006).

\bibitem{97} P. E. Bosted et al., Phys. Rev. $\bf{C75}$, 035203
(2007).

\bibitem{98} Y. Prok et al., Phys. Lett. $\bf{B672}$, 12 (2009).

\bibitem{99} E.D. Bloom and E.J. Gilman, Phys. Rev. Lett. $\bf{25}$, 1140 (1970).

\bibitem{100} E.D. Bloom and E.J. Gilman, Phys.Rev. $\bf{D4}$, 2901 (1971).

\bibitem{101} R.Petronzio, S.Simula, G.Ricco, Phys. Rev.$\bf{D67}$, 094004
(2003), hep-ph/0301206.

\end{thebibliography}
\end{document}